\documentclass[aps,prd,twocolumn,showkeys,amsmath,amssymb]{revtex4}
\usepackage{graphicx}
\usepackage{epstopdf}
\usepackage{multirow}
\usepackage{subfigure}
\usepackage[colorlinks,citecolor=blue,anchorcolor=red,menucolor=red,linkcolor=red,filecolor=red,runcolor=red,urlcolor=blue,frenchlinks=red]{hyperref}
\allowdisplaybreaks[3]

\begin{document}

\title{Excited $\Xi_c^0$ baryons within the QCD rum rule approach}

\author{Hui-Min Yang$^1$}
\author{Hua-Xing Chen$^{2}$}
\email{hxchen@seu.edu.cn}
\author{Qiang Mao$^{3}$}

\affiliation{
$^1$School of Physics, Beihang University, Beijing 100191, China\\
$^2$School of Physics, Southeast University, Nanjing 210094, China\\
$^3$Department of Electrical and Electronic Engineering, Suzhou University, Suzhou 234000, China}

\begin{abstract}
We systematically study mass spectra and decay properties of $P$-wave $\Xi_c^\prime$ baryons of the $SU(3)$ flavor $\mathbf{6}_F$, using the methods of QCD sum rules and light-cone sum rules within the framework of heavy quark effective theory. Our results suggest that the three excited $\Xi_c^0$ baryons recently observed by LHCb can be well explained as $P$-wave $\Xi_c^\prime$ baryons: the $\Xi_c(2923)^0$ and $\Xi_c(2939)^0$ are partner states of $J^P = 1/2^-$ and $3/2^-$ respectively, both of which contain one $\lambda$-mode orbital excitation; the $\Xi_c(2965)^0$ has $J^P = 3/2^-$, and also contains one $\lambda$-mode orbital excitation. We propose to search for another $P$-wave $\Xi_c^\prime$ state of $J^P = 5/2^-$ in the $\Lambda_c K/\Xi_c \pi$ mass spectral in future experiments. Its mass is about $56^{+30}_{-35}$~MeV larger than the $\Xi_c(2965)^0$, and its width is about $18.1^{+19.7}_{-~8.3}$~MeV.
\end{abstract}

\maketitle

\section{Introduction}

The light quarks and gluons circle around the nearly static heavy quark inside the heavy baryon. This system is the QCD analogue of the hydrogen, but bounded by the strong interaction~\cite{Korner:1994nh,Manohar:2000dt,Klempt:2009pi}. In three recent LHCb experiments~\cite{Aaij:2017nav,Aaij:2020cex,Aaij:2020yyt} its spectra are found to have beautiful fine structures: five excited $\Omega_c^0$ baryons were observed in the experiment~\cite{Aaij:2017nav}; four excited $\Omega_b^-$ baryons were observed in the experiment~\cite{Aaij:2020cex}; in the very recent experiment~\cite{Aaij:2020yyt} three excited $\Xi_c^0$ baryons were observed simultaneously in the $\Lambda^+_c K^-$ mass spectrum, whose parameters were measured to be:
\begin{eqnarray}
\nonumber \Xi_c(2923)^0   &:& M = 2923.04 \pm 0.25 \pm 0.20 \pm 0.14~{\rm MeV} \, ,
\\                 && \Gamma = 7.1 \pm 0.8 \pm 1.8~{\rm MeV} \, ,
\\ \nonumber \Xi_c(2939)^0   &:& M = 2938.55 \pm 0.21 \pm 0.17 \pm 0.14~{\rm MeV} \, ,
\\                 && \Gamma = 10.2 \pm 0.8 \pm 1.1~{\rm MeV} \, ,
\\ \nonumber \Xi_c(2965)^0   &:& M = 2964.88 \pm 0.26 \pm 0.14 \pm 0.14~{\rm MeV} \, ,
\\                 && \Gamma = 14.1 \pm 0.9 \pm 1.3~{\rm MeV} \, .
\end{eqnarray}
These excited $\Omega_c^0/\Omega_b^-/\Xi_c^0$ baryons are good candidates of $P$-wave charmed and bottom baryons, whose observations have proved the rich internal structure of (heavy) hadrons~\cite{Chen:2016spr,Copley:1979wj,pdg}.

The LHCb Collaboration~\cite{Aaij:2020yyt} further pointed out that the $\Xi_c(2923)^0$ and $\Xi_c(2939)^0$ baryons are probably the sub-structures of $\Xi_c(2930)^0$~\cite{Aubert:2007eb,Li:2017uvv}, while the $\Xi_c(2965)^0$ and $\Xi_c(2970)^0$~\cite{Chistov:2006zj} might be different states. Various phenomenological methods and models have been applied to study these baryons. In Ref.~\cite{Lu:2020ivo} the author uses the constituent quark model to interpret the $\Xi_c(2923)^0$ and $\Xi_c(2939)^0$ as the $\lambda$-mode $P$-wave $\Xi_c^\prime$ baryons of $J^P = 3/2^-$ and $5/2^-$, and the $\Xi_c(2965)^0$ as the $J^P = 1/2^+$ $\Xi_c^\prime(2S)$ state. In Ref.~\cite{Wang:2020gkn} the authors use the chiral quark model to interpret them as the $\lambda$-mode $P$-wave $\Xi_c^\prime$ baryons of $J^P = 3/2^-$, $3/2^-$, and $5/2^-$, respectively. In Ref.~\cite{Agaev:2020fut} the authors use the method of QCD sum rules to interpret the $\Xi_c(2923)^0$ and $\Xi_c(2939)^0$ as $P$-wave $\Xi_c^\prime$ baryons of $J^P = 1/2^-$ and $3/2^-$, and the $\Xi_c(2965)^0$ as the $J^P = 1/2^+$ $\Xi_c^\prime(2S)$ or $\Xi_c(2S)$ state. In Ref.~\cite{Zhu:2020jke} the authors use the molecular picture to interpret the $\Xi_c(2923)^0$ as a $D\bar \Lambda$-$D\bar \Sigma$ molecule.

Besides, many phenomenological methods and models have been applied to understand the $\Xi_c(2930)^0$ and $\Xi_c(2970)^0$ previously observed by BaBar~\cite{Aubert:2007eb} and Belle~\cite{Chistov:2006zj}, such as
various quark models~\cite{Capstick:1986bm,Chen:2007xf,Ebert:2007nw,Garcilazo:2007eh,Roberts:2007ni,Zhong:2007gp,Valcarce:2008dr,Ebert:2011kk,Ortega:2012cx,Yoshida:2015tia,Nagahiro:2016nsx,Wang:2017kfr,Ye:2017yvl},
various molecular explanations~\cite{GarciaRecio:2008dp,Liang:2014eba,Yu:2018yxl,Nieves:2019jhp,Huang:2017dwn},
the chiral perturbation theory~\cite{Lu:2014ina,Cheng:2015naa},
Lattice QCD~\cite{Padmanath:2013bla,Padmanath:2017lng,Bahtiyar:2020uuj},
and QCD sum rules~\cite{Bagan:1991sg,Groote:1996em,Huang:2000tn,Zhang:2008pm,Aliev:2009jt,Wang:2010it,Aliev:2018ube}, etc.
We refer to the reviews~\cite{Chen:2016spr,Cheng:2015iom,Crede:2013kia,Amhis:2019ckw} and references therein for detailed discussions.

We have systematically studied mass spectra and decay properties of $P$-wave heavy baryons in Refs.~\cite{Chen:2015kpa,Chen:2017sci,Yang:2019cvw,Yang:2020zrh} using the methods of QCD sum rules~\cite{Shifman:1978bx,Reinders:1984sr} and light-cone sum rules~\cite{Balitsky:1989ry,Braun:1988qv,Chernyak:1990ag,Ball:1998je,Ball:2006wn} within the framework of heavy quark effective theory (HQET)~\cite{Grinstein:1990mj,Eichten:1989zv,Falk:1990yz}. The results were combined in Ref.~\cite{Yang:2020zrh} so that a rather complete study within HQET was performed on both mass spectra and decay properties of $P$-wave bottom baryons. There we predicted four $\Xi^\prime_b$ baryons, three of which have finite and limited widths, while the rest one has a (nearly) zero width:
\begin{eqnarray}
\nonumber [\Xi_b^\prime(1/2^-), 1, 1, \lambda]   &:& M = 6.21 \pm 0.11~{\rm GeV} \, ,
\\                 && \Gamma = 4.7~{^{+5.8}_{-3.3}}~{\rm MeV} \, ,
\\ \nonumber [\Xi_b^\prime(3/2^-), 1, 1, \lambda]   &:& M = 6.22 \pm 0.11~{\rm MeV} \, ,
\\                 && \Gamma = 1.8~{^{+1.1}_{-1.0}}~{\rm MeV} \, ,
\\ \nonumber [\Xi_b^\prime(3/2^-), 2, 1, \lambda]   &:& M = 6.23 \pm 0.15~{\rm GeV} \, ,
\\                 && \Gamma = 27.3~{^{+28.5}_{-14.2}}~{\rm MeV} \, ,
\\ \nonumber [\Xi_b^\prime(5/2^-), 2, 1, \lambda]   &:& M = 6.24 \pm 0.14~{\rm MeV} \, ,
\\                 && \Gamma = 12.7~{^{+12.4}_{-~6.1}}~{\rm MeV} \, .
\end{eqnarray}
Their mass splittings are:
\begin{eqnarray}
M_{[\Xi_b^\prime(3/2^-), 1, 1, \lambda]} - M_{[\Xi_b^\prime(1/2^-), 1, 1, \lambda]} &=& 7 \pm 2~{\rm MeV} \, ,
\\ \nonumber M_{[\Xi_b^\prime(5/2^-), 2, 1, \lambda]} - M_{[\Xi_b^\prime(3/2^-), 2, 1, \lambda]} &=& 11 \pm 5~{\rm MeV} \, .
\end{eqnarray}
The above notations will be explained later, and we refer to Ref.~\cite{Yang:2020zrh} for their detailed decay channels.
From our previous results~\cite{Yang:2020zrh}, we guess that the $\Xi_c(2923)^0$, $\Xi_c(2939)^0$, and $\Xi_c(2965)^0$ are just the charmed partners of the $[\Xi_b^\prime(1/2^-), 1, 1, \lambda]$, $[\Xi_b^\prime(3/2^-), 1, 1, \lambda]$, and $[\Xi_b^\prime(3/2^-), 2, 1, \lambda]$, respectively. To verify this, in this paper we follow the same approach used in Refs.~\cite{Chen:2015kpa,Chen:2017sci,Yang:2019cvw,Yang:2020zrh} to study the above excited $\Xi_c^0$ baryons recently observed by LHCb~\cite{Aaij:2020cex}. We shall find that all of them can be interpreted as $P$-wave $\Xi_c^\prime$ baryons of the $SU(3)$ flavor $\mathbf{6}_F$, so that both their mass spectra and decay properties can be well explained.

This paper is organized as follows. In Sec.~\ref{sec:mass} we introduce our notations for $P$-wave $\Xi_c^\prime$ baryons, and categorize them into four charmed baryon multiplets $[\Xi_c^\prime, 1, 0, \rho]$, $[\Xi_c^\prime, 0, 1, \lambda]$, $[\Xi_c^\prime, 1, 1, \lambda]$, and $[\Xi_c^\prime, 2, 1, \lambda]$. We use them to perform QCD sum rule analyses within the framework of heavy quark effective theory, and calculate their masses. Then in Sec.~\ref{sec:decay} we study their decay properties, including their $S$-wave and $D$-wave decays into ground-state charmed baryons and pseudoscalar mesons ($\pi$ or $K$) as well as their $S$-wave decays into ground-state charmed baryons and vector mesons ($\rho$ or $K^*$). In Sec.~\ref{sec:summary} we discuss the results and conclude this paper.

\section{Mass spectra from QCD sum rules}
\label{sec:mass}

\begin{figure*}[tb]
\begin{center}
\scalebox{0.7}{\includegraphics{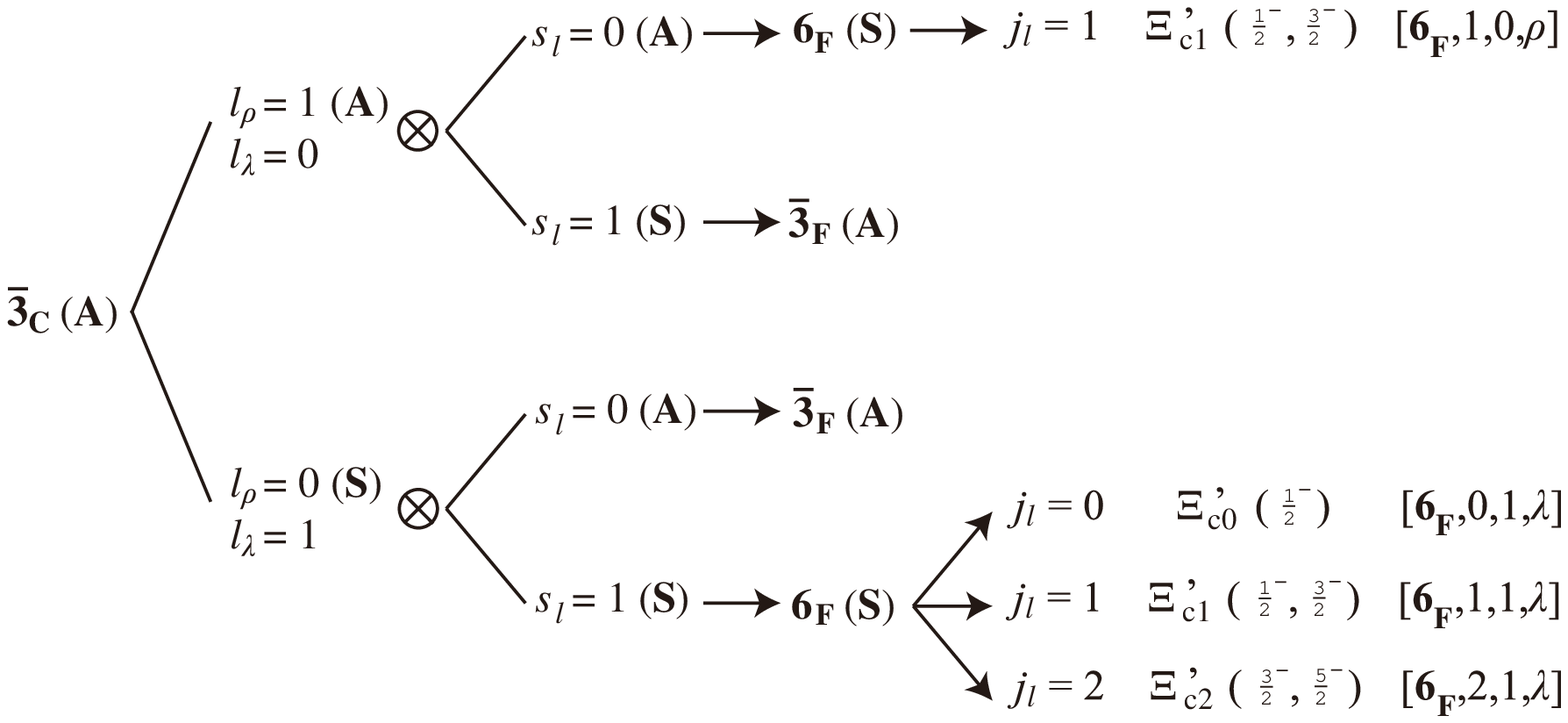}}
\caption{Categorization of $P$-wave $\Xi_c^\prime$ baryons.
\label{fig:pwave}}
\end{center}
\end{figure*}

We follow Ref.~\cite{Chen:2007xf} and use the same notations to describe $P$-wave $\Xi_c^\prime$ baryons of the $SU(3)$ flavor $\mathbf{6}_F$. Each baryon consists of one charm quark and two light quarks, and contains one orbital excitation, which can be either between the two light quarks ($l_\rho = 1$) or between the charm quark and the two-light-quark system ($l_\lambda = 1$). Hence, there are $\rho$-mode excited $\Xi_c^\prime$ baryons ($l_\rho = 1$ and $l_\lambda = 0$) and $\lambda$-mode ones ($l_\rho = 0$ and $l_\lambda = 1$). Together with the color, flavor, and spin degrees of freedom, its internal structures are:
\begin{itemize}

\item Color structure of the two light quarks is antisymmetric, that is the color $\mathbf{\bar 3}_C$.

\item Flavor structure of the two light quarks is symmetric, that is the $SU(3)$ flavor $\mathbf{6}_F$.

\item Spin structure of the two light quarks can be either antisymmetric ($s_l \equiv s_{qq} = 0$) or symmetric ($s_l = 1$).

\item Orbital structure of the two light quarks can be either antisymmetric ($l_\rho = 1$) or symmetric ($l_\rho = 0$).

\end{itemize}
Considering that the total structure of the two light quarks is antisymmetric due to the Pauli principle, we can categorize $P$-wave $\Xi_c^\prime$ baryons into four multiplets, denoted as $[\mathbf{6}_F, j_l, s_l, \rho/\lambda]$. We show them in Fig.~\ref{fig:pwave}, where $j_l$ denotes the total angular momentum of the light components ($j_l = s_l \otimes l_\rho \otimes l_\lambda$). Every multiplet contains one or two $\Xi_c^\prime$ baryons, whose total angular momenta are $j = j_l \otimes s_c = | j_l \pm 1/2 |$, with $s_c$ the charm quark spin.

To avoid confusions, we add a note here that the symbol $\Xi_c$ but not $\Xi_c^\prime$ was used in the previous section to describe the baryon states observed in the LHCb experiment~\cite{Aaij:2020yyt}. This is because the flavor symmetry of the two light quarks can not be differentiated so that $\Xi_c$ and $\Xi_c^\prime$ can not be differentiated neither in the experiment. However, theoretically this can be done, and we use $\Xi_c$ and $\Xi_c^\prime$ to denote baryons belonging to the $SU(3)$ flavor $\mathbf{\bar 3}_F$ and $\mathbf{6}_F$, respectively.

We have systematically studied mass spectra of $P$-wave charmed baryons in Ref.~\cite{Chen:2015kpa} using QCD sum rules within HQET. In this method we calculate the baryon mass through
\begin{equation}
m_{\Xi_c^\prime(j^P),j_l,s_l,\rho/\lambda} = m_c + \overline{\Lambda}_{\Xi_c^\prime,j_l,s_l,\rho/\lambda} + \delta m_{\Xi_c^\prime(j^P),j_l,s_l,\rho/\lambda} \, ,
\label{eq:mass}
\end{equation}
where $m_c$ is the charm quark mass, $\overline{\Lambda}_{\Xi_c^\prime,j_l,s_l,\rho/\lambda} = \overline{\Lambda}_{\Xi_c^\prime(|j_l-1/2|),j_l,s_l,\rho/\lambda} = \overline{\Lambda}_{\Xi_c^\prime(j_l+1/2),j_l,s_l,\rho/\lambda}$ is extracted from the mass sum rules at the leading order, and $\delta m_{\Xi_c^\prime(j^P),j_l,s_l,\rho/\lambda}$ is extracted from the mass sum rules at the ${\mathcal O}(1/m_c)$ order. We refer to Ref.~\cite{Chen:2015kpa} for their explicit expressions.

Eq.~(\ref{eq:mass}) tells that the $\Xi_c^\prime$ mass depends significantly on the charm quark mass. Hence, there exists considerable (theoretical) uncertainty in our results for absolute values of baryon masses, and we can not distinguish the three excited $\Xi_c^0$ baryons observed by LHCb~\cite{Aaij:2020yyt} only by using their mass spectra. However, the mass splittings within the same multiplets are produced at the ${\mathcal O}(1/m_c)$ order with much less uncertainty, giving more useful information.

We can extract even more valuable information from decay properties of $P$-wave $\Xi_c^\prime$ baryons. Before doing this we fine-tune one of the two free parameters in mass sum rules, the threshold value $\omega_c$, to get a better description of the LHCb experiment~\cite{Aaij:2020yyt}. The other free parameter, the Borel mass $T$, can be determined by using two criteria: the first is to require the high-order power corrections to be less than 30\%, and the second is to require the pole contribution of $P$-wave $\Xi_c^\prime$ baryons to be larger than 20\%. Note that there may exist some lower state, making it not easy to get an idea pole contribution at 50\%. This is quite similar to the QCD sum rule study on excited heavy mesons~\cite{Zhou:2014ytp}. Besides, the small pole contribution is mathematically due to the large powers of $s$ in the spectral function, which makes the suppression of the Borel transformation on the continuum not so effective. For example, see Ref.~\cite{Chen:2014vha} where the pole contribution of the $d^*(2380)$ is only about 0.0002 due to the large power of $s$ in its spectral function. As a compensation, we need the third criterion, which requires the mass dependence on the threshold value $\omega_c$ to be weak.

After fixing these two parameters, all the other parameters can be calculated using the method of QCD sum rules with HQET. We summarize the obtained results in Table~\ref{tab:mass}, together with the parameters that are necessary to calculate decay widths through light-cone sum rules. Note that they are slightly different from those parameters for excited bottom baryons~\cite{Yang:2020zrh}. Their uncertainties are due to various QCD sum rule parameters~\cite{Yang:1993bp,Narison:2002pw,Gimenez:2005nt,Jamin:2002ev,Ioffe:2002be,Ovchinnikov:1988gk,Ellis:1996xc,pdg}:
%
%%%%%%%%%%%%%%%%%%%%%%%%%%%%%%%%%%%%%%%%%%%%%%%%%%%%%%%%%%%%%%%%%%%%%%%%%%%%%%
\begin{eqnarray}
\nonumber m_s(1\mbox{ GeV}) &=& (137 \pm 27) \mbox{ MeV} \, ,
\\ \nonumber  \langle g_s^2GG\rangle &=& (0.48\pm 0.14) \mbox{ GeV}^4 \, ,
\\ \nonumber  \langle\bar qq \rangle &=& -(0.24 \pm 0.01)^3 \mbox{ GeV}^3 \, ,
\\ \langle\bar ss \rangle &=& (0.8\pm 0.1)\times \langle\bar qq \rangle \, ,
\label{condensates}
\\
\nonumber \langle g_s\bar q\sigma G q\rangle &=& - M_0^2\times\langle\bar qq\rangle \, ,
\\
\nonumber \langle g_s\bar s\sigma G s\rangle &=& - M_0^2\times\langle\bar ss\rangle \, ,
\\
\nonumber M_0^2 &=& (0.8 \pm 0.2) \mbox{ GeV}^2 \, .
\end{eqnarray}
%%%%%%%%%%%%%%%%%%%%%%%%%%%%%%%%%%%%%%%%%%%%%%%%%%%%%%%%%%%%%%%%%%%%%%%%%%%%%%
%

Our results suggest that the $\rho$-mode excitation is lower than the $\lambda$-mode, a behaviour which is consistent with our previous results for their corresponding $SU(3)$ flavor $\mathbf{\bar 3}_F$ multiplets~\cite{Chen:2015kpa}, but in contrast to the quark model expectation~\cite{Yoshida:2015tia}. However, this is possible simply because that the mass differences between different multiplets have considerable uncertainties in our QCD sum rule calculations, similar to the absolute values of baryon masses, but different from the mass differences within the same multiplet.

\begin{table*}[hbt]
\begin{center}
\renewcommand{\arraystretch}{1.35}
\caption{Mass spectra of $P$-wave $\Xi_c^\prime$ baryons belonging to the $SU(3)$ flavor $\mathbf{6}_F$ representation, evaluated using QCD sum rules within HQET. Here we also list the parameters that are necessary to calculate their decay widths through light-cone sum rules.}
\begin{tabular}{c | c  c  c | c  c | c | c}
\hline\hline
\multirow{2}{*}{~~~~~Multiplets~~~~~} & ~~~~$\omega_c$~~~~ & ~~~Working region~~~ & ~~~~~~~~~$\overline{\Lambda}$~~~~~~~~~ & ~~~Baryon~~~ & ~~~~~Mass~~~~~ & ~~Difference~~ & ~~~~~~~~~~$f$~~~~~~~~~~
\\                                    &        (GeV)       &         (GeV)        &                 (GeV)                  &    ($j^P$)   &      (GeV)     &       (MeV)    &      (GeV$^{4}$)
\\ \hline\hline
\multirow{2}{*}{$[\mathbf{6}_F(\Xi_c^\prime), 1, 0, \rho]$}
& \multirow{2}{*}{1.87} & \multirow{2}{*}{$0.26< T < 0.34$} & \multirow{2}{*}{$1.36^{+0.12}_{-0.08}$} & $\Xi_c^\prime(1/2^-)$ & $2.88^{+0.15}_{-0.13}$ & \multirow{2}{*}{$13^{+6}_{-5}$} & $0.059^{+0.017}_{-0.011}$
\\ \cline{5-6}\cline{8-8}
& & & & $\Xi_c^\prime(3/2^-)$ & $2.89^{+0.15}_{-0.13}$ & &$0.028^{+0.008}_{-0.005}$
\\ \hline
$[\mathbf{6}_F(\Xi_c^\prime), 0, 1, \lambda]$
& 1.57 & $0.27< T < 0.29$ & $1.22^{+0.08}_{-0.07} $ & $\Xi_c^\prime(1/2^-)$ & $2.90^{+0.13}_{-0.12} $ & -- & $0.041^{+0.010}_{-0.008}$
\\ \hline
\multirow{2}{*}{$[\mathbf{6}_F(\Xi_c^\prime), 1, 1, \lambda]$}
& \multirow{2}{*}{1.72} & \multirow{2}{*}{$T=0.34 $} & \multirow{2}{*}{$1.14^{+0.09}_{-0.08}$} & $\Xi_c^\prime(1/2^-)$ & $2.91^{+0.13}_{-0.12}$ & \multirow{2}{*}{$38^{+15}_{-13}$} & $0.041^{+0.008}_{-0.007}$
\\ \cline{5-6}\cline{8-8}
& & & & $\Xi_c^\prime(3/2^-)$ & $2.95^{+0.12}_{-0.11}$ & &$0.019^{+0.004}_{-0.003}$
\\ \hline
\multirow{2}{*}{$[\mathbf{6}_F(\Xi_c^\prime), 2, 1, \lambda]$}
& \multirow{2}{*}{1.72} & \multirow{2}{*}{$0.27< T < 0.32$} & \multirow{2}{*}{$1.24^{+0.15}_{-0.09}$} & $\Xi_c^\prime(3/2^-)$ & $2.96^{+0.24}_{-0.15}$ & \multirow{2}{*}{$66^{+29}_{-25}$} & $0.057^{+0.020}_{-0.012}$
\\ \cline{5-6}\cline{8-8}
& & & & $\Xi_c^\prime(5/2^-)$ & $3.02^{+0.23}_{-0.14}$ & &$0.034^{+0.012}_{-0.007}$
\\ \hline \hline
\end{tabular}
\label{tab:mass}
\end{center}
\end{table*}

\section{Widths from light-cone sum rules}
\label{sec:decay}

We have systematically studied decay properties of $P$-wave heavy baryons in Refs.~\cite{Chen:2017sci,Yang:2019cvw,Yang:2020zrh} using light-cone sum rules within HQET, and the results are combined in Ref.~\cite{Yang:2020zrh} to study $P$-wave bottom baryons. In the present study we replace the bottom quark by the charm quark, and redo all the calculations. We summarized all the sum rule equations in Appendix~\ref{app:sumrules}. Their extracted results are given in Table~\ref{tab:width}, where we have investigated all the possible $S$-wave and $D$-wave decays of $P$-wave $\Xi_c^\prime$ baryons into ground-state charmed baryons and light pseudoscalar mesons.

Their uncertainties are due to the parameters given in Table~\ref{tab:mass} as well as various light-cone sum rule parameters~\cite{Ball:1998je,Ball:2006wn}. Because there are many input parameters with uncertainties (some of them are given in Appendix~\ref{app:parameter}), the uncertainties of our results are not so small, {\it i.e.}, they can be as large as $X^{+200\%}_{-~67\%}$. Especially, the parameter $a_2^{\pi/K}$ of $\phi_{\pi/K;2}$ is $0.25 \pm 0.15$~\cite{Ball:1998je,Ball:2006wn}, which causes the major uncertainties.

During the calculations, we have used the following mass values:
\begin{itemize}

\item For the $[\mathbf{6}_F(\Xi_c^{\prime}), 1, 0, \rho]$ doublet, we use the following mass values taken from their mass sum rules:
\begin{eqnarray}
M_{[\Xi_c^{\prime}(1/2^-), 1, 0, \rho]} &=& 2.88^{+0.15}_{-0.13}~{\rm GeV} \, ,
\\ \nonumber M_{[\Xi_c^{\prime}(3/2^-), 1, 0, \rho]} &=& 2.89^{+0.15}_{-0.13}~{\rm GeV} \, .
\end{eqnarray}

\item For the $[\mathbf{6}_F(\Xi_c^{\prime}), 0, 1, \lambda]$ singlet, we use the following mass value taken from its mass sum rules:
\begin{eqnarray}
M_{[\Xi_c^{\prime}(1/2^-), 0, 1, \lambda]} &=& 2.90^{+0.13}_{-0.12}~{\rm GeV}\, .
\end{eqnarray}

\item For the $[\mathbf{6}_F(\Xi_c^{\prime}), 1, 1, \lambda]$ doublet, we use the masses of $\Xi_c(2923)^0$ and $\Xi_c(2939)^0$ measured by LHCb~\cite{Aaij:2020yyt}:
\begin{eqnarray}
M_{[\Xi_c^{\prime}(1/2^-), 1, 1, \lambda]} &=& M_{\Xi_c(2923)^0} = 2923.04~{\rm GeV} \, ,
\\ \nonumber M_{[\Xi_c^{\prime}(3/2^-), 1, 1, \lambda]} &=& M_{\Xi_c(2939)^0} = 2938.55~{\rm GeV} \, .
\end{eqnarray}

\item For the $[\mathbf{6}_F(\Xi_c^{\prime}), 2, 1, \lambda]$ doublet, we use the mass of $\Xi_c(2965)^0$ as well as their mass sum rules:
\begin{eqnarray}
M_{[\Xi_c^{\prime}(3/2^-), 2, 1, \lambda]} &=& M_{\Xi_c(2965)^0} = 2964.88~{\rm MeV}\, ,
\\ \nonumber M_{[\Xi_c^{\prime}(5/2^-), 2, 1, \lambda]} &=& M_{[\Xi_c^{\prime}(3/2^-), 2, 1, \lambda]} + 56~{\rm MeV} \, .
\end{eqnarray}

\end{itemize}

From Table~\ref{tab:width} we quickly find that the $\Xi_c(2923)^0$, $\Xi_c(2939)^0$, and $\Xi_c(2965)^0$ may be interpreted as the $P$-wave $\Xi_c^\prime$ baryons $[\Xi_c^{\prime}(1/2^-), 1, 1, \lambda]$, $[\Xi_c^{\prime}(3/2^-), 1, 1, \lambda]$, and $[\Xi_c^{\prime}(3/2^-), 2, 1, \lambda]$, respectively. However, there exist three discrepancies between our theoretical results and the LHCb measurements~\cite{Aaij:2020yyt}: a) the missing of the $\Lambda_c K$ decay channel for the former two baryons, b) the mass splitting between the former two baryons, and c) the total widths of the latter two baryons.

\begin{table*}[hbtp]
\begin{center}
\caption{Decay properties of $P$-wave $\Xi_c^\prime$ baryons belonging to the $SU(3)$ flavor $\mathbf{6}_F$ representation. Here the baryons are categorized according to the heavy quark effective theory (HQET). In the fifth column $\Gamma_S$ and $\Gamma_D$ denote the relevant decay channel to be $S$-wave and $D$-wave, respectively. Their possible experimental candidates are given in the last column for comparisons. Note that there exists considerable uncertainty in our results for absolute values of baryon masses (the third column), but the mass splittings within the same doublets (the fourth column) are produced quite well with much less uncertainty.}
\renewcommand{\arraystretch}{1.3}
\begin{tabular}{   c | c | c | c | c | c | c}
\hline\hline
  \multirow{2}{*}{~Multiplet~} & ~~Baryon~~ & ~~~~Mass~~~~ & Difference & \multirow{2}{*}{~~~~~~~~~~~~~~~~~~~~~Decay channel~~~~~~~~~~~~~~~~~~~~~}& Total width  & \multirow{2}{*}{Candidate}
\\  & ($j^P$) & ({GeV}) & ({MeV}) & & ({MeV}) &
\\ \hline\hline
\multirow{4}{*}{$[\mathbf{6}_F, 1, 0, \rho]$} & $\Xi_c^{\prime}({1/2}^-)$ & $2.88^{+0.15}_{-0.13}$& \multirow{6}{*}{$13^{+6}_{-5}$} &
%%%%%%%%%%%%%%%%%%%%%%%%%%%%%%%%%%%%%%%%%%%%%%%%%%%%%%%%%%
$\begin{array}{l}
\Gamma_S\left(\Xi_c^{\prime}({1/2}^-)\to\Xi_c^{\prime}\pi\right)=110^{+170}_{-~80}~{\rm MeV}\\
\Gamma_D\left(\Xi_c^{\prime}({1/2}^-)\to \Xi_c^{*}\pi\right)=0.15^{+0.23}_{-0.11}~{\rm MeV}\\
\Gamma_S\left(\Xi_c^{\prime}({1/2}^-)\to\Xi_c\rho\to\Xi_c\pi\pi\right)=2 \cdot 10^{-4}~{\rm MeV}\\
\Gamma_S\left(\Xi_c^{\prime}({1/2}^-)\to\Xi_c^{\prime}\rho\to\Xi_c^{\prime}\pi\pi\right)=5 \cdot 10^{-9}~{\rm MeV}
\end{array}$ &$110^{+170}_{-~80}$&--
%%%%%%%%%%%%%%%%%%%%%%%%%%%%%%%%%%%%%%%%%%%%%%%%%%%%%%%%%%
\\ \cline{2-3}\cline{5-7}
                                                                           & $\Xi_c^{\prime}({3/2}^-)$ & $2.89^{+0.15}_{-0.13}$ &&
$\begin{array}{l}
\Gamma_D\left(\Xi_c^{\prime}({3/2}^-)\to \Xi_c^{\prime}\pi\right)=0.63^{+0.99}_{-0.45}~{\rm MeV} \\
\Gamma_S\left(\Xi_c^{\prime}({3/2}^-)\to \Xi_c^{*}\pi\right)=58^{+88}_{-39}~{\rm MeV}\\
\Gamma_D\left(\Xi_c^{\prime}({3/2}^-)\to \Xi_c^{*}\pi\right)=0.03^{+0.05}_{-0.02}~{\rm MeV}\\
\Gamma_S\left(\Xi_c^{\prime}({3/2}^-)\to\Xi_c\rho\to\Xi_c\pi\pi\right)=5 \cdot 10^{-4}~{\rm MeV}\\
\Gamma_S\left(\Xi_c^{\prime}({3/2}^-)\to\Xi_c^{\prime}\rho\to\Xi_c^{\prime}\pi\pi\right)=2 \cdot 10^{-9}~{\rm MeV}
\end{array}$  &$58.5^{+88.4}_{-39.4}$&--
\\ \cline{1-7}
$[\mathbf{6}_F,0,1,\lambda]$&$\Xi_c^{\prime}({1/2}^-)$&$2.90^{+0.13}_{-0.12}$& -- &
$\begin{array}{l}
\Gamma_S\left(\Xi_c^{\prime}({1/2}^-)\to\Lambda_c K\right)=400^{+610}_{-270}~{\rm MeV}\\
\Gamma_S\left(\Xi_c^{\prime}({1/2}^-)\to \Xi_c \pi\right)=360^{+550}_{-250}~{\rm MeV}\\
\Gamma_S\left(\Xi_c^{\prime}({1/2}^-)\to\Xi_c^{\prime}\rho\to\Xi_c^{\prime}\pi\pi\right)=0.03~{\rm MeV}
\end{array}$
&$760^{+820}_{-370}$&--
\\ \cline{1-7}
\multirow{4}{*}{$[\mathbf{6}_F,1,1,\lambda]$}&$\Xi^\prime_c({1/2}^-)$&$2.91^{+0.13}_{-0.12}$&\multirow{6}{*}{$38^{+15}_{-13}$}&
$\begin{array}{l}
\Gamma_S\left(\Xi_c^{\prime}({1/2}^-)\to \Xi_c^{\prime}\pi\right)=11.7^{+15.0}_{-~8.0}~{\rm MeV}\\
\Gamma_D\left(\Xi_c^{\prime}({1/2}^-)\to\Xi_c^{*}\pi\right)=0.12^{+0.22}_{-0.10}~{\rm MeV}\\
\Gamma_S\left(\Xi_c^{\prime}({1/2}^-)\to\Lambda_c K^{*}\to\Lambda_c K\pi\right)=4 \cdot 10^{-8}~{\rm MeV}\\
\Gamma_S\left(\Xi_c^{\prime}({1/2}^-)\to\Xi_c\rho\to\Xi_c\pi\pi\right)=1.7^{+7.6}_{-1.7}~{\rm MeV}\\
\Gamma_S\left(\Xi_c^{\prime}({1/2}^-)\to\Xi_c^{\prime}\rho\to\Xi_c^{\prime}\pi\pi\right)=0.38^{+0.54}_{-0.30}~{\rm MeV}\\
\Gamma_S\left(\Xi_c^{\prime}({1/2}^-)\to\Xi_c^{*}\rho\to\Xi_c^{*}\pi\pi\right)=2 \cdot 10^{-7}~{\rm MeV}
\end{array}$&$13.9^{+16.8}_{-~8.2}$&$\Xi_c(2923)^0$
\\ \cline{2-3}\cline{5-7}
&$\Xi_c^{\prime}({3/2}^-)$&$2.95^{+0.12}_{-0.11}$&&
$\begin{array}{l}
\Gamma_D\left(\Xi_c^{\prime}({3/2}^-)\to\Xi_ c^{\prime}\pi\right)=0.67^{+1.06}_{-0.52}~{\rm MeV}\\
\Gamma_S\left(\Xi_c^{\prime}({3/2}^-)\to \Xi_c^{*}\pi\right)=3.3^{+4.3}_{-2.3}~{\rm MeV}\\
\Gamma_D\left(\Xi_c^{\prime}({3/2}^-)\to\Xi_c^{*}\pi\right)=0.05^{+0.08}_{-0.04}~{\rm MeV}\\
\Gamma_S\left(\Xi_c^{\prime}({3/2}^-)\to\Lambda_c K^{*}\to\Lambda_c K\pi\right)=2 \cdot 10^{-4}~{\rm MeV}\\
\Gamma_S\left(\Xi_c^{\prime}({3/2}^-)\to\Xi_c\rho\to\Xi_c\pi\pi\right)=0.21^{+0.60}_{-0.20}~{\rm MeV}\\
\Gamma_S\left(\Xi_c^{\prime}({3/2}^-)\to\Xi_c^{\prime}\rho\to\Xi_c^{\prime}\pi\pi\right)=0.12^{+0.19}_{-0.10}~{\rm MeV}\\
\Gamma_S\left(\Xi_c^{\prime}({3/2}^-)\to\Xi_c^{*}\rho\to\Xi_c^{*}\pi\pi\right)=1 \cdot 10^{-3}~{\rm MeV}
\end{array}$&$4.4^{+4.5}_{-2.3}$
& \multirow{8}{*}{$\begin{array}{c} \Xi_c(2939)^0 \\ and \\ \Xi_c(2965)^0\end{array}$}
\\ \cline{1-6}
\multirow{3}{*}{$[\mathbf{6}_F,2,1,\lambda]$}&$\Xi_c^{\prime}({3/2}^-)$&$2.96^{+0.24}_{-0.15}$&\multirow{8}{*}{$66^{+29}_{-25}$}&
$\begin{array}{l}
\Gamma_D\left(\Xi_c^{\prime}({3/2}^-)\to \Lambda_c K\right)=9.8^{+17.9}_{-~7.2}~{\rm MeV}\\
\Gamma_D\left(\Xi_c^{\prime}({3/2}^-)\to \Xi_c\pi\right)=17.0^{+29.7}_{-12.0}~{\rm MeV}\\
\Gamma_D\left(\Xi_c^{\prime}({3/2}^-)\to \Sigma_c K\right)=0.003^{+0.015}_{-0.003}~{\rm MeV}\\
\Gamma_D\left(\Xi_c^{\prime}({3/2}^-)\to \Xi_c^{\prime}\pi\right)=2.3^{+4.0}_{-1.7}~{\rm MeV}\\
\Gamma_S\left(\Xi_c^{\prime}({3/2}^-)\to \Xi_c^{*}\pi\right)= 2 \cdot 10^{-4}~{\rm MeV}\\
\Gamma_D\left(\Xi_c^{\prime}({3/2}^-)\to \Xi_c^{*}\pi\right)=0.19^{+0.33}_{-0.14}~{\rm MeV}\\
\Gamma_S\left(\Xi_c^{\prime}({3/2}^-)\to\Xi_c^{\prime}\rho\to\Xi_c^{\prime}\pi\pi\right)=1.4^{+2.2}_{-1.0}~{\rm MeV}\\
\Gamma_S\left(\Xi_c^{\prime}({3/2}^-)\to\Xi_c^{*}\rho\to\Xi_c^{*}\pi\pi\right)=1 \cdot 10^{-3}~{\rm MeV}
\end{array}$&$30.7^{+35.0}_{-14.2}$&
\\ \cline{2-3} \cline{5-7}
&$\Xi^\prime_c({5/2}^-)$&$3.02^{+0.23}_{-0.14}$&&
$\begin{array}{l}
\Gamma_D\left(\Xi_c^{\prime}({5/3}^-)\to \Lambda_c K\right)=6.3^{+11.4}_{-~4.6}~{\rm MeV}\\
\Gamma_D\left(\Xi_c^{\prime}({5/2}^-)\to \Xi_c \pi\right)=9.6^{+15.8}_{-~6.8}~{\rm MeV}\\
\Gamma_D\left(\Xi_c^{\prime}({5/2}^-)\to \Sigma_c K\right)=0.02^{+0.09}_{-0.02}~{\rm MeV}\\
\Gamma_D\left(\Xi_c^{\prime}({5/2}^-)\to \Xi_c^{\prime} \pi\right)=0.70^{+1.30}_{-0.54}~{\rm MeV}\\
\Gamma_D\left(\Xi_c^{\prime}({5/2}^-)\to \Sigma_c^{*} \pi\right)=4 \cdot 10^{-3}~{\rm MeV}\\
\Gamma_D\left(\Xi_c^{\prime}({5/2}^-)\to \Xi_c^{*} \pi\right)=1.5^{+2.6}_{-1.1}~{\rm MeV}\\
\Gamma_S\left(\Xi_c^{\prime}({5/2}^-)\to\Xi_c^{*}\rho\to\Xi_c^{*}\pi\pi\right)=0.02~{\rm MeV}
\end{array}$&$18.1^{+19.7}_{-~8.3}$&--
\\ \hline \hline
\end{tabular}
\label{tab:width}
\end{center}
\end{table*}

\begin{table*}[hbtp]
\begin{center}
\caption{Decay properties of $P$-wave $\Xi_c^\prime$ baryons of the $SU(3)$ flavor $\mathbf{6}_F$. The first column lists the baryons categorized according to the heavy quark effective theory (HQET), and the third column lists the baryons after considering the mixing. The possible experimental candidates are given in the last column for comparisons.}
\renewcommand{\arraystretch}{1.3}
\scalebox{0.93}{\begin{tabular}{   c|c | c | c | c | c | c | c}
\hline\hline
  \multirow{2}{*}{HQET state}&\multirow{2}{*}{Mixing}&\multirow{2}{*}{Mixed state} & Mass & Difference & ~~~~~~~~~~~~~~~~~~Decay channel~~~~~~~~~~~~~~~~~~& Width  & \multirow{2}{*}{Candidate}
\\  &&&   ({GeV}) & ({MeV}) & ({MeV})& ({MeV}) &
\\ \hline\hline
$[\Xi_c^\prime({1\over2}^-),0,1,\lambda]$&\multirow{7}{*}{$\theta_1 \approx 0^\circ$}&$[\Xi_c^{\prime}({1\over2}^-),0,1,\lambda]$&$2.90^{+0.13}_{-0.12}$& -- &
$\begin{array}{l}
\Gamma_S\left(\Xi_c^{\prime}({1/2}^-)\to\Lambda_c K\right)=400^{+610}_{-270}\\
\Gamma_S\left(\Xi_c^{\prime}({1/2}^-)\to \Xi_c \pi\right)=360^{+550}_{-250}\\
\Gamma_S\left(\Xi_c^{\prime}({1/2}^-)\to\Xi_c^{\prime}\rho\to\Xi_c^{\prime}\pi\pi\right)=0.03
\end{array}$
&$760^{+820}_{-370}$&--
\\ \cline{1-1}\cline{3-8}
$[\Xi_c^\prime({1\over2}^-),1,1,\lambda]$&&$[\Xi_c^\prime({1\over2}^-),1,1,\lambda]$&$2.91^{+0.13}_{-0.12}$&\multirow{6}{*}{$27^{+16}_{-27}$}&
$\begin{array}{l}
\Gamma_S\left(\Xi_c^{\prime}({1/2}^-)\to \Xi_c^{\prime}\pi\right)=11.7^{+15.0}_{-~8.0}\\
\Gamma_D\left(\Xi_c^{\prime}({1/2}^-)\to\Xi_c^{*}\pi\right)=0.12^{+0.22}_{-0.10}\\
\Gamma_S\left(\Xi_c^{\prime}({1/2}^-)\to\Lambda_c K^{*}\to\Lambda_c K\pi\right)=4 \cdot 10^{-8}\\
\Gamma_S\left(\Xi_c^{\prime}({1/2}^-)\to\Xi_c\rho\to\Xi_c\pi\pi\right)=1.7^{+7.6}_{-1.7}\\
\Gamma_S\left(\Xi_c^{\prime}({1/2}^-)\to\Xi_c^{\prime}\rho\to\Xi_c^{\prime}\pi\pi\right)=0.38^{+0.54}_{-0.30}\\
\Gamma_S\left(\Xi_c^{\prime}({1/2}^-)\to\Xi_c^{*}\rho\to\Xi_c^{*}\pi\pi\right)=2 \cdot 10^{-7}
\end{array}$&$13.9^{+16.8}_{-~8.2}$&$\Xi_c(2923)^0$
\\ \cline{1-4}\cline{6-8}
$[\Xi_c^{\prime}({3\over2}^-),1,1,\lambda]$&\multirow{10}{*}{$\theta_2={37\pm5^\circ}$}&$\Xi_c^\prime({3\over2}^-)_1$&$2.94^{+0.12}_{-0.11}$&&
$\begin{array}{l}
\Gamma_D\left(\Xi_c^{\prime}({3/2}^-)\to \Lambda_c K\right)=2.3^{+4.3}_{-1.7}\\
\Gamma_D\left(\Xi_c^{\prime}({3/2}^-)\to \Xi_c\pi\right)=4.6^{+8.1}_{-3.3}\\
\Gamma_D\left(\Xi_c^{\prime}({3/2}^-)\to\Xi_ c^{\prime}\pi\right)=2.0^{+2.2}_{-1.2}\\
\Gamma_S\left(\Xi_c^{\prime}({3/2}^-)\to \Xi_c^{*}\pi\right)=2.1^{+2.6}_{-1.5}\\
\Gamma_D\left(\Xi_c^{\prime}({3/2}^-)\to\Xi_c^{*}\pi\right)=0.14^{+0.16}_{-0.08}\\
\Gamma_S\left(\Xi_c^{\prime}({3/2}^-)\to\Lambda_c K^{*}\to\Lambda_c K\pi\right)=2 \cdot 10^{-6}\\
\Gamma_S\left(\Xi_c^{\prime}({3/2}^-)\to\Xi_c\rho\to\Xi_c\pi\pi\right)=0.13^{+0.39}_{-0.13}\\
\Gamma_S\left(\Xi_c^{\prime}({3/2}^-)\to\Xi_c^{\prime}\rho\to\Xi_c^{\prime}\pi\pi\right)=0.5^{+0.6}_{-0.3}\\
\Gamma_S\left(\Xi_c^{\prime}({3/2}^-)\to\Xi_c^{*}\rho\to\Xi_c^{*}\pi\pi\right)=1 \cdot 10^{-3}
\end{array}$&$11.8^{+9.8}_{-4.2}$
& $\Xi_c(2939)^0 $
\\ \cline{1-1}\cline{3-8}
$[\Xi_c^\prime({3\over2}^-),2,1,\lambda]$&&$\Xi_c^\prime({3\over2}^-)_2$&$2.97^{+0.24}_{-0.15}$&\multirow{3}{*}{$56^{+30}_{-35}$}&
$\begin{array}{l}
\Gamma_D\left(\Xi_c^{\prime}({3/2}^-)\to \Lambda_c K\right)=6.3^{+11.6}_{-~4.7}\\
\Gamma_D\left(\Xi_c^{\prime}({3/2}^-)\to \Xi_c\pi\right)=10.9^{+19.1}_{-~7.8}\\
\Gamma_D\left(\Xi_c^{\prime}({3/2}^-)\to \Sigma_c K\right)=4 \cdot 10^{-4}\\
\Gamma_D\left(\Xi_c^{\prime}({3/2}^-)\to \Xi_c^{\prime}\pi\right)=0.37^{+1.89}_{-0.36}\\
\Gamma_S\left(\Xi_c^{\prime}({3/2}^-)\to \Xi_c^{*}\pi\right)= 1.3^{+1.80}_{-0.94}\\
\Gamma_D\left(\Xi_c^{\prime}({3/2}^-)\to \Xi_c^{*}\pi\right)=0.03^{+0.16}_{-0.03}\\
\Gamma_S\left(\Xi_c^{\prime}({3/2}^-)\to\Lambda_c K^{*}\to\Lambda_c K\pi\right)=2 \cdot 10^{-5}\\
\Gamma_S\left(\Xi_c^{\prime}({3/2}^-)\to\Xi_c\rho\to\Xi_c\pi\pi\right)=0.12^{+0.36}_{-0.12}\\
\Gamma_S\left(\Xi_c^{\prime}({3/2}^-)\to\Xi_c^{\prime}\rho\to\Xi_c^{\prime}\pi\pi\right)=0.37^{+1.15}_{-0.36}\\
\Gamma_S\left(\Xi_c^{\prime}({3/2}^-)\to\Xi_c^{*}\rho\to\Xi_c^{*}\pi\pi\right)=5 \cdot 10^{-3}
\end{array}$&$19.4^{+22.5}_{-~9.1}$&$\Xi_c(2965)^0$
\\ \cline{1-4} \cline{6-8}
$[\Xi^\prime_c({5\over2}^-),2,1,\lambda]$&--&$[\Xi^\prime_c({5\over2}^-),2,1,\lambda]$&$3.02^{+0.23}_{-0.14}$&&
$\begin{array}{l}
\Gamma_D\left(\Xi_v^{\prime}({5/3}^-)\to \Lambda_c K\right)=6.3^{+11.4}_{-~4.6}\\
\Gamma_D\left(\Xi_c^{\prime}({5/2}^-)\to \Xi_c \pi\right)=9.6^{+15.8}_{-~6.8}\\
\Gamma_D\left(\Xi_c^{\prime}({5/2}^-)\to \Sigma_c K\right)=0.02^{+0.09}_{-0.02}\\
\Gamma_D\left(\Xi_c^{\prime}({5/2}^-)\to \Xi_c^{\prime} \pi\right)=0.70^{+1.30}_{-0.54}\\
\Gamma_D\left(\Xi_c^{\prime}({5/2}^-)\to \Sigma_c^{*} \pi\right)=4 \cdot 10^{-3}\\
\Gamma_D\left(\Xi_c^{\prime}({5/2}^-)\to \Xi_c^{*} \pi\right)=1.5^{+2.6}_{-1.1}\\
\Gamma_S\left(\Xi_c^{\prime}({5/2}^-)\to\Xi_c^{*}\rho\to\Xi_c^{*}\pi\pi\right)=0.02
\end{array}$&$18.1^{+19.7}_{-~8.3}$
&--
\\ \hline \hline
\end{tabular}}
\label{tab:result}
\end{center}
\end{table*}

These discrepancies are possible and reasonable because the HQET is an effective theory, which works well for bottom baryons but not so well for charmed baryons. Hence, the three $J=1/2^-$ $\Xi_c^\prime$ baryons can mix together and the three $J=3/2^-$ $\Xi_c^\prime$ baryons can also mix together. Especially, a tiny mixing angle $\theta_1 \approx 0^\circ$ is enough to make it possible to observe all of them in the $\Lambda_c K$ decay channel.

Since the two $\Xi_c^\prime(3/2^-)$ baryons belonging to the $[{\bf 6}_F, 1, 1, \lambda]$ and $[{\bf 6}_F, 2, 1, \lambda]$ doublets are very close to each other, in this paper we consider their mixing explicitly:
\begin{eqnarray}
\left(\begin{array}{c}
|\Xi_c^\prime(3/2^-)\rangle_1\\
|\Xi_c^\prime(3/2^-)\rangle_2
\end{array}\right)
&=&
\left(\begin{array}{cc}
\cos\theta_2 & \sin\theta_2 \\
-\sin\theta_2 & \cos\theta_2
\end{array}\right)
\\ \nonumber && ~~~~~~~~~~ \times
\left(\begin{array}{c}
|\Xi_c^\prime(3/2^-),1,1,\lambda\rangle\\
|\Xi_c^\prime(3/2^-),2,1,\lambda\rangle
\end{array}\right) \, ,
\end{eqnarray}
where $\theta_2$ is the mixing angle. Fine-tuning it to be $\theta_2 = 37\pm5^\circ$, we show the obtained results in Table~\ref{tab:result}. Very quickly, we find that this mixing mediates the widths of $[\Xi_c^\prime(3/2^-), 1, 1, \lambda]$ and $[\Xi_c^\prime(3/2^-), 2, 1, \lambda]$, and decreases the mass splitting within the $[{\bf 6}_F, 1, 1, \lambda]$ doublet:
\begin{eqnarray*}
\nonumber M_{[\Xi_c^{\prime}(3/2^-), 1, 1, \lambda]} &:& 2.95^{+0.12}_{-0.11}~{\rm GeV} \longrightarrow 2.94^{+0.12}_{-0.11}~{\rm GeV} \, ,
\\ \nonumber \Gamma_{[\Xi_c^{\prime}(3/2^-), 1, 1, \lambda]} &:& 4.4^{+4.5}_{-2.3}~{\rm MeV} \longrightarrow 11.8^{+9.8}_{-4.2}~{\rm MeV} \, ,
\\ M_{[\Xi_c^{\prime}(3/2^-), 2, 1, \lambda]} &:& 2.96^{+0.24}_{-0.15}~{\rm GeV} \longrightarrow 2.97^{+0.24}_{-0.15}~{\rm GeV} \, ,
\\ \nonumber \Gamma_{[\Xi_c^{\prime}(3/2^-), 2, 1, \lambda]} &:& 30.7^{+35.0}_{-14.2}~{\rm MeV} \longrightarrow 19.4^{+22.5}_{-~9.1}~{\rm MeV} \, ,
\\ \nonumber \Delta M_{[\Xi_c^{\prime}, 1, 1, \lambda]} &:& 38^{+15}_{-13}~{\rm MeV} \longrightarrow 27^{+16}_{-27}~{\rm MeV} \, ,
\\ \nonumber \Delta M_{[\Xi_c^{\prime}, 2, 1, \lambda]} &:& 66^{+29}_{-25}~{\rm MeV} \longrightarrow 56^{+30}_{-35}~{\rm MeV} \, .
\end{eqnarray*}
Now the $\Xi_c(2939)^0$ and $\Xi_c(2965)^0$ can be well explained by using the two $J^P = 3/2^-$ baryons $|\Xi_c^\prime(3/2^-)\rangle_1$ and $|\Xi_c^\prime(3/2^-)\rangle_2$, respectively.

Our QCD sum rule results are similar to the quark model calculations~\cite{Wang:2020gkn}. Besides, it is interesting to compare our results with the $SU(3)$ flavor symmetry. Take $[\Xi_c^\prime(3/2^-),2,1,\lambda]$ as an example, its partial decay widths to the $\Lambda_c K$ and $\Xi_c \pi$ final states are evaluated to be $9.8^{+17.9}_{-~7.2}$~MeV and $17.0^{+29.7}_{-12.0}$~MeV, respectively. Their ratio is about 0.58, similar to following factor derived from the $SU(3)$ flavor symmetry:
\begin{eqnarray}
\nonumber \mathcal{R} \equiv {\Gamma(\Xi_c^\prime \to \Lambda_c K) \over \Gamma(\Xi_c^\prime \to \Xi_c \pi)} &\approx& {\Gamma(\Xi_c^{\prime0} \to \Lambda_c^+ K^-) \over 1.5 \times \Gamma(\Xi_c^{\prime0} \to \Xi_c^+ \pi^-)}
\\ \nonumber &\sim& {g_{\Xi_c^{\prime0} \to \Lambda_c^+ K^-}^2 \over 1.5 \times g_{\Xi_c^{\prime0} \to \Xi_c^+ \pi^-}^2}
\\ &=& 0.67 \, .
\end{eqnarray}
If considering that $M_{\Lambda_c} + M_K = 2782$~MeV is larger than $M_{\Xi_c} + M_\pi = 2607$~MeV, we can understand the above diversity even better.

Our results are obtained using light-cone sum rules within HQET. It is also interesting to compare them with the results of Ref.~\cite{Agaev:2020fut}, which are obtained using full QCD light-cone sum rule method. There the widths of the $1P$ $\Xi_c^{\prime}$ baryons with $J^P = 1/2^-$ and $3/2^-$ are calculated to be $7.2 \pm 1.4$~MeV and $10.1 \pm 2.1$ MeV respectively, supporting the interpretation of $\Xi_c(2939)^0$ and $\Xi_c(2965)^0$ as such states. These two values are consistent with our results that $\Gamma_{[\Xi_c^{\prime}(1/2^-), 1, 1, \lambda]} = 13.9^{+16.8}_{-~8.2}$~MeV and $\Gamma_{[\Xi_c^{\prime}(3/2^-), 1, 1, \lambda]} = 11.8^{+9.8}_{-4.2}$~MeV, and our interpretations are the same for the $\Xi_c(2939)^0$ and $\Xi_c(2965)^0$.

\section{Summary and Discussions}
\label{sec:summary}

In the present study we have investigated $P$-wave $\Xi_c^\prime$ baryons of the $SU(3)$ flavor $\mathbf{6}_F$ by systematically studying their mass spectra and decay properties using the methods of QCD sum rules and light-cone sum rules within the framework of heavy quark effective theory. The obtained results are summarized in Tables~\ref{tab:width} and \ref{tab:result}, from which we can well understand the three excited $\Xi_c^0$ baryons recently observed by LHCb~\cite{Aaij:2020yyt} as $P$-wave $\Xi_c^\prime$ baryons of the $SU(3)$ flavor $\mathbf{6}_F$.

There can be as many as seven $P$-wave $\Xi_c^\prime$ baryons, belonging to four multiplets:
\begin{eqnarray}
\nonumber \Xi_c^\prime(1/2^-) \, , \Xi_c^\prime(3/2^-)~~~~~~~~~~~~~~ &\in& [{\bf 6}_F, 1, 0, \rho] \, ,
\\ \nonumber \Xi_c^\prime(1/2^-) ~~~~~~~~~~~~~~~~~~~~~~~~~~~~          &\in& [{\bf 6}_F, 0, 1, \lambda] \, ,
\\ \nonumber \Xi_c^\prime(1/2^-) \, , \Xi_c^\prime(3/2^-)~~~~~~~~~~~~~~ &\in& [{\bf 6}_F, 1, 1, \lambda] \, ,
\\ \nonumber \Xi_c^\prime(3/2^-) \, , \Xi_c^\prime(5/2^-) &\in& [{\bf 6}_F, 2, 1, \lambda] \, .
\end{eqnarray}
Our results suggest:
\begin{itemize}

\item The $\Xi_c(2923)^0$ and $\Xi_c(2939)^0$ can be interpreted as the $P$-wave $\Xi_c^\prime$ baryons of $J^P = 1/2^-$ and $3/2^-$ respectively, both of which belong to the $[{\bf 6}_F(\Xi_c^\prime), 1, 1, \lambda]$ doublet. The $\Xi_c(2965)^0$ can be interpreted as the $P$-wave $\Xi_c^\prime$ baryon of $J^P = 3/2^-$, belonging to the $[{\bf 6}_F(\Xi_c^\prime), 2, 1, \lambda]$ doublet. It has a partner state, $\Xi_c^\prime$ of $J^P = 5/2^-$, whose mass is about $56^{+30}_{-35}$~MeV larger and width about $18.1^{+19.7}_{-~8.3}$~MeV. We propose to search for it in the $\Lambda_c K/\Xi_c \pi$ mass spectral in future experiments.

\item The HQET is an effective theory, which works well for bottom baryons but not so well for charmed baryons. This suggests that the three $J=1/2^-$ $\Xi_c^\prime$ baryons can mix together and the three $J=3/2^-$ ones can also mix together, making it possible to observe all of them in the $\Lambda_c K$ invariant mass spectrum. Especially, in this paper we have explicitly considered the mixing between the two $\Xi_c^\prime(3/2^-)$ baryons belonging to the $[{\bf 6}_F, 1, 1, \lambda]$ and $[{\bf 6}_F, 2, 1, \lambda]$ doublets, which mediates their widths as well as decreases the mass splitting within the $[{\bf 6}_F, 1, 1, \lambda]$ doublet. The obtained results can be used to better describe the LHCb experiment~\cite{Aaij:2020yyt}.

\item The width of $[\Xi_c^0(1/2^-), 0, 1, \lambda]$ is too large for it to be observed in experiments. The widths of $\Xi_c^\prime(1/2^-)$ and $\Xi_c^\prime(3/2^-)$ belonging to the $[{\bf 6}_F(\Xi_c^\prime), 1, 0, \rho]$ doublet are evaluated to be about 110~MeV and 59~MeV respectively, making them not so easy to be observed. We notice that there is ``an additional component'' observed by LHCb in the energy region around 2900~MeV~\cite{Aaij:2020yyt}, which may be due to these two states.

\end{itemize}
The above conclusions are obtained by combining our systematical studies on mass spectra, mass splittings within the same multiplets, and decay properties of $P$-wave $\Xi_c^\prime$ baryons. Moreover, we have taken into account the five excited $\Omega_c^0$ and four excited $\Omega_b^-$ baryons observed by LHCb~\cite{Aaij:2017nav,Aaij:2020cex}, whose correspondences may be~\cite{Chen:2020mpy}:
\begin{eqnarray}
\nonumber [{\bf 6}_F(?/2^-), 1, 0, \rho] &:& \Xi_c^\prime(?/2^-) \sim \Omega_c^0(3000) \sim \Omega_b^-(6316) \, ,
\\
\nonumber [{\bf 6}_F(1/2^-), 1, 1, \lambda] &:& \Xi_c^0(2923)\, \sim \Omega_c^0(3050) \sim \Omega_b^-(6330) \, ,
\\
\nonumber [{\bf 6}_F(3/2^-), 1, 1, \lambda] &:& \Xi_c^0(2939)\, \sim \Omega_c^0(3066) \sim \Omega_b^-(6340) \, ,
\\
\nonumber [{\bf 6}_F(3/2^-), 2, 1, \lambda] &:& \Xi_c^0(2965)\, \sim \Omega_c^0(3090) \sim \Omega_b^-(6350) \, ,
\\
\nonumber [{\bf 6}_F(5/2^-), 2, 1, \lambda] &:& \Xi_c^\prime(5/2^-) \sim \Omega_c^0(3119) \sim \Omega_b^-(5/2^-) .
\end{eqnarray}
We shall detailedly discuss this in our future work~\cite{pwave}.

In the present study we have investigated $P$-wave $\Xi_c^{\prime}$ baryons within the heavy quark effective theory. Because the finite charm quark mass breaks this symmetry explicitly, we have also considered the mixing effect between baryons having the same spin-parity quantum number, such as the mixing between $[\Xi_c^{\prime}({3\over2}^-),1,1,\lambda]$ and $[\Xi_c^{\prime}({3\over2}^-),2,1,\lambda]$. One can study them within some other schemes. For example, in Refs.~\cite{Liang:2020hbo,Wang:2020gkn} the authors studied excited $\Xi_c^{\prime}$ and $\Omega_b$ baryons in the j-j coupling scheme based on the chiral quark model. However, there they also considered the mixing effect, so the obtained ``physical'' basis can be eventually the same.

To end this paper, we note that the conclusions of the present study are just possible explanations, and there exist some other possibilities for the three excited $\Xi_c^0$ baryons observed by LHCb~\cite{Aaij:2020yyt}. Further experimental and theoretical studies are still demanded to fully understand them, since the beautiful fine structures of the excited singly heavy baryons observed in the three LHCb experiments~\cite{Aaij:2017nav,Aaij:2020cex,Aaij:2020yyt} have proved the rich internal structure of (heavy) hadrons, and their relevant studies are significantly improving our knowledge of the strong interaction.

\section*{Acknowledgments}

We thank Qi-Fang L{\"u} for useful discussions.
This project is supported by
the National Natural Science Foundation of China under Grants No.~11722540 and No.~12075019,
the Fundamental Research Funds for the Central Universities,
and
the Foundation for Young Talents in College of Anhui Province (Grant No.~gxyq2018103).

\appendix

%
%=====================================================================================
%=====================================================================================
\section{Input parameters}
\label{app:parameter}
%=====================================================================================
%=====================================================================================
%

We list masses of ground-state charmed baryons used in this paper~\cite{pdg}:
\begin{eqnarray}
   \nonumber        \Lambda_{c}(1/2^+)  ~:~ m&=&2286.46 \mbox{ MeV} \, ,
\\ \nonumber        \Xi_{c}(1/2^+)  ~:~ m&=&2469.34 \mbox{ MeV} \, ,
\\ \nonumber        \Sigma_{c}(1/2^+)    ~:~ m&=&2453.54 \mbox{ MeV} \, ,
\\ \nonumber        \Sigma_{c}^{*}(3/2^+)    ~:~ m&=&2518.1 \mbox{ MeV} \, , \,
\\ \nonumber        \Xi_{c}^{\prime}(1/2^+)  ~:~ m&=&2576.8 \mbox{ MeV} \, ,
\\ \nonumber        \Xi_{b}^{*}(3/2^+)  ~:~ m&=&2645.9 \mbox{ MeV} \, , \,
\\ \nonumber        \Omega_b(1/2^+)  ~:~ m&=&2695.2 \mbox{ MeV}\, ,
\\ \nonumber        \Omega_b^*(3/2^+)  ~:~ m&=&2765.9 \mbox{ MeV} \, .
\end{eqnarray}
Their sum rule parameters can be found in Refs.~\cite{Liu:2007fg,Chen:2017sci}.

We list masses and decay widths of pseudoscalar and vector mesons used in this paper~\cite{pdg}:
\begin{eqnarray}
 \pi(0^-) ~:~ m&=& 138.04 {\rm~MeV} \, ,
\\ \nonumber K(0^-) ~:~ m&=& 495.65 {\rm~MeV} \, ,
\\ \nonumber \rho(1^-) ~:~ m&=& 775.21 {\rm~MeV} \, ,\,
\\ \nonumber                                \Gamma&=& 148.2 {\rm~MeV} \, ,\,  {g}_{\rho \pi \pi} = 5.94 \, ,
\\  \nonumber         K^*(1^-) ~:~ m&=& 893.57 {\rm~MeV} \, ,\,
\\                                 \Gamma&=& 49.1 {\rm~MeV} \, ,\,  {g}_{K^* K \pi} = 3.20 \, ,
\end{eqnarray}
where the two coupling constants ${g}_{\rho \pi \pi}$ and ${g}_{K^* K \pi}$ are evaluated using the experimental widths of the $\rho$ and $K^*$~\cite{pdg} through the Lagrangians:
\begin{eqnarray}
\nonumber \mathcal{L}_{\rho \pi \pi} &=& {g}_{\rho \pi \pi} \times \left( \rho_\mu^0 \pi^+ \partial^\mu \pi^- - \rho_\mu^0 \pi^- \partial^\mu \pi^+ \right) + \cdots \, ,
\\ \nonumber \mathcal{L}_{K^* K \pi} &=& {g}_{K^* K \pi} K^{*+}_\mu \times \left( K^- \partial^\mu \pi^0 - \partial^\mu K^- \pi^0 \right) + \cdots \, .
\\
\end{eqnarray}

We also list the light-cone distribution amplitudes of the $K$ meson, which are taken from Ref.~\cite{Ball:2006wn}. We refer to Refs.~\cite{Ball:1998je,Ball:2006wn} for detailed discussions. The light-cone distribution amplitudes of the $K$ meson used in this paper are:
\begin{widetext}
\begin{eqnarray}
\langle 0 | \bar q(z)\gamma_\mu\gamma_5 s(-z) | K(q) \rangle
&=& i f_K q_\mu \int_0^1 du \, e^{i (2u-1) q \cdot z} \left(\phi_{2;K}(u) + \frac{1}{4}\,z^2 \phi_{4;K}(u)\right)
\\ \nonumber && ~~~~~~~~~~ + \frac{i}{2}\, f_K\, \frac{1}{q \cdot z}\, z_\mu  \int_0^1 du \, e^{i (2u-1) q \cdot z} \psi_{4;K}(u) \, ,
\\ \langle 0 | \bar q(z) i\gamma_5 s(-z) | K(q) \rangle
&=& \frac{f_K m_K^2}{m_s+m_q}\, \int_0^1 du \, e^{i(2u-1) q \cdot z} \, \phi^{p}_{3;K}(u) \, ,
\\ \langle 0 | \bar q(z) \sigma_{\alpha\beta}\gamma_5 s(-z) | K(q) \rangle &=& -\frac{i}{3}\, \frac{f_K m_K^2}{m_s+m_q}  (q_\alpha z_\beta- q_\beta z_\alpha) \int_0^1 du \, e^{i(2u-1) q \cdot z}\,\phi^{\sigma}_{3;K}(u) \, ,
\\ \langle 0 | \bar q(z)\gamma_\mu\gamma_5 gG_{\alpha\beta}(vz)s(-z)|K(q)\rangle
&=& q_\mu (q_\alpha z_\beta - q_\beta z_\alpha)\, \frac{1}{q \cdot z}\, f_{K} \Phi_{4;K}(v,q \cdot z)
\\ \nonumber && ~~~~~~~~~~ + (q_\beta g_{\alpha\mu}^\perp - q_\alpha g_{\beta\mu}^\perp) f_{K} \Psi_{4;K}(v,q \cdot z) \, ,
\\ \langle 0 | \bar q(z)\gamma_\mu i g\widetilde{G}_{\alpha\beta}(vz)s(-z)| K(q)\rangle\
&=& q_\mu (q_\alpha z_\beta - q_\beta z_\alpha)\, \frac{1}{q \cdot z}\, f_K \widetilde\Phi_{4;K}(v,q \cdot z)
\\ \nonumber && ~~~~~~~~~~ + (q_\beta g_{\alpha\mu}^\perp - q_\alpha g_{\beta\mu}^\perp) f_{K} \widetilde\Psi_{4;K}(v,q \cdot z) \, ,
\\ \langle 0 | \bar q(z) \sigma_{\mu\nu}\gamma_5 gG_{\alpha\beta}(vz) s(-z)| K(q)\rangle
&=& i\,f_{3K} \left(q_\alpha q_\mu g_{\nu\beta}^\perp - q_\alpha q_\nu g_{\mu\beta}^\perp - (\alpha\leftrightarrow\beta) \right) \times
\\ \nonumber && ~~~~~~~~~~ \times \int {\cal D}\underline{\alpha} \, e^{-iq \cdot z(\alpha_2 -\alpha_1 + v\alpha_3)} {\Phi}_{3;K}(\alpha_1,\alpha_2,\alpha_3) \, ,
\end{eqnarray}
where $\widetilde{G}_{\mu\nu} = \frac{1}{2}\epsilon_{\mu\nu \rho\sigma} G^{\rho\sigma}$.

\section{Sum rule equations}
\label{app:sumrules}

In this appendix we give the sum rule equations used to study $S$-wave and $D$-wave decays of $P$-wave $\Xi_c^{\prime}$ baryons into ground-state charmed baryons and pseudoscalar mesons.

The sum rule equations for the $\Xi_b^{\prime-}[{1\over2}^-]$ belonging to $[\mathbf{6}_F, 1 , 0, \rho]$ are
\begin{eqnarray}
&&G_{\Xi_c^{\prime0}[{1\over2}^-]\to\Xi_c^{*+}\pi^-}^D(\omega,\omega^{\prime})= \frac{g_{\Xi_c^{\prime0}[{1\over2}^-]\to\Xi_c^{*+}\pi^- } f_{\Xi_c^{\prime0}[{1\over2}^-]}f_{\Xi_c^{*+}}}{(\bar{\Lambda}_{\Xi_c^{\prime0}[{1\over2}^-]}-\omega^{\prime})(\bar{\Lambda}_{\Xi_c^{*+}}-\omega)}
\\ \nonumber&=& \int_0^\infty dt \int_0^1 du e^{i(1-u)\omega^\prime t} e^{iu\omega t}\times 4\times \Big(\frac{f_{\pi}m_s u}{128\pi^2}\phi_{4;\pi}(u)+\frac{f_{\pi} u}{24}\langle\bar s s\rangle \phi_{2;\pi}(u)
\\ \nonumber &&+\frac{f_{\pi} m_s u}{8\pi^2 t^2}\phi_{2;\pi}(u)+\frac{f_{\pi} m_{\pi}^2 u}{24 (m_u+m_d)\pi^2 t^2}\phi_{3;\pi}^{\sigma}(u)+\frac{f_{\pi} m_s u t^2}{576 (m_u+m_d)}\langle\bar s s\rangle \phi_{3;\pi}^{\sigma}(u)+\frac{f_{\pi} u t^2}{384}\langle\bar s s\rangle \phi_{4;\pi}(u)
\\ \nonumber &&+\frac{f_{\pi} u t^2}{384}\langle g_s \bar s \sigma G s\rangle \phi_{2;\pi}(u)+\frac{f_{\pi} u t^4}{6144}\langle g_s \bar s \sigma G s\rangle \phi_{4;\pi}(u)\Big)
\\ \nonumber &-&\int_0^\infty dt \int_0^1 du \int \mathcal{D}\underline{\alpha}e^{i\omega^\prime t(\alpha_2+u\alpha_3)}e^{i\omega t(1-\alpha_2-u\alpha_3)}\times{1\over2}\times\Big(-\frac{i f_{3\pi} u v\cdot q}{4\pi^2 t}\Phi_{3;\pi}(\underline{\alpha})+\frac{i f_{3\pi} \alpha_2 u v\cdot q}{4 \pi^2 t}\Phi_{3;\pi}(\underline{\alpha})
\\ \nonumber &&+\frac{i f_{3\pi}\alpha_3 u v\cdot q}{4\pi^2 t}\Phi_{3;\pi}(\underline{\alpha})-\frac{i f_{3\pi} u v\cdot q}{4\pi^2 t}\Phi_{3;\pi}(\underline{\alpha})+\frac{i f_{3\pi}\alpha_2 v\cdot q}{4\pi^2 t}\Phi_{3\pi}(\underline{\alpha})-\frac{i f_{3\pi} v\cdot q}{4\pi^2 t}\Phi_{3;\pi}(\underline{\alpha})+\frac{f_{3\pi} u}{4\pi^2 t^2}\Phi_{3;\pi}(\underline{\alpha})\Big) \, ,
\\
&&G_{\Xi_c^{\prime0}[{1\over2}^-]\to\Xi_c^{\prime+}\pi^-}^S(\omega,\omega^{\prime})= \frac{g_{\Xi_c^{\prime0}[{1\over2}^-]\to\Xi_c^{\prime+}\pi^- } f_{\Xi_c^{\prime0}[{1\over2}^-]}f_{\Xi_c^{\prime+}}}{(\bar{\Lambda}_{\Xi_c^{\prime0}[{1\over2}^-]}-\omega^{\prime})(\bar{\Lambda}_{\Xi_c^{\prime+}}-\omega)}
\\ \nonumber&=& \int_0^\infty dt \int_0^1 du e^{i(1-u)\omega^\prime t} e^{iu\omega t}\times 4\times \Big(\frac{3 f_{\pi} m_{\pi}^2}{4\pi^2 (m_u+m_d) t^4}\phi_{3;\pi}^p(u)+\frac{i f_{\pi} m_{\pi}^2 v\cdot q}{8\pi^2 (m_u+m_d) t^3}\phi_{3;\pi}^{\sigma}(u)
\\ \nonumber &&-\frac{3 i f_{\pi} m_s}{16 \pi^2 t^3 v\cdot q}\psi_{4;\pi}(u)-\frac{i f_{\pi}}{16 t v\cdot q}\langle\bar s s\rangle\psi_{4;\pi}(u)+\frac{f_{\pi} m_s m_{\pi}^2}{32 (m_u+m_d)}\phi_{3;\pi}^p(u)-\frac{i f_{\pi} t}{256 v\cdot q}\langle g_s \bar s \sigma G s\rangle \psi_{4;\pi}(u)
\\ \nonumber &&+\frac{i f_{\pi} m_s m_{\pi}^2 t v\cdot q}{192(m_u+m_d)}\langle\bar s s\rangle\phi_{3;\pi}^{\sigma}(u)\Big) \, .
\end{eqnarray}
The sum rule equations for the $\Xi_b^{\prime-}[{3\over2}^-]$ belonging to $[\mathbf{6}_F, 1 , 0, \rho]$ are
\begin{eqnarray}
&&G_{\Xi_c^{\prime0}[{3\over2}^-]\to\Xi_c^{\prime+}\pi^-}^D(\omega,\omega^{\prime})= \frac{g_{\Xi_c^{\prime0}[{3\over2}^-]\to\Xi_c^{\prime+}\pi^- } f_{\Xi_c^{\prime0}[{3\over2}^-]}f_{\Xi_c^{\prime+}}}{(\bar{\Lambda}_{\Xi_c^{\prime0}[{3\over2}^-]}-\omega^{\prime})(\bar{\Lambda}_{\Xi_c^{\prime+}}-\omega)}
\\ \nonumber &=& \int_0^\infty dt \int_0^1 du e^{i(1-u)\omega^\prime t} e^{iu\omega t}\times 4\times \Big(\frac{f_{\pi} m_s u}{8\pi^2 t^2}\phi_{2;\pi}(u)+\frac{f_{\pi} m_{\pi}^2 u}{24 (m_u+m_d)\pi^2 t^2}\phi_{3;\pi}^{\sigma}(u)
\\ \nonumber &&+\frac{f_{\pi} m_s u}{128 \pi^2}\phi_{4;\pi}(u)+\frac{f_{\pi} u}{24}\langle\bar s s\rangle \phi_{2;\pi}(u)+\frac{f_{\pi} m_s  m_{\pi}^2 u t^2}{576 (m_u+m_d)}\langle\bar s s\rangle\phi_{3;\phi}^{\sigma}(u)+\frac{f_{\pi} u t^2}{384}\langle\bar s s\rangle\phi_{4;\pi}(u)
\\ \nonumber &&+\frac{f_{\pi} u t^2}{384}\langle g_s \bar s\sigma G s\rangle\phi_{2;\pi}(u)+\frac{f_{\pi} u t^4}{6144}\langle g_s \bar s\sigma G s\rangle\phi_{4;\pi}(u)\Big)
\\ \nonumber &-&\int_0^\infty dt \int_0^1 du \int \mathcal{D}\underline{\alpha}e^{i\omega^\prime t(\alpha_2+u\alpha_3)}e^{i\omega t(1-\alpha_2-u\alpha_3)}\times{1\over2}\times\Big(-\frac{f_{3\pi} u}{4\pi^2 t^2}\Phi_{3;\pi}(\underline{\alpha})+\frac{if_{3\pi}\alpha_3 u^2 v\cdot q}{4\pi^2 t}\Phi_{3;\pi}(\underline{\alpha})
\\ \nonumber &&+\frac{if_{3\pi}\alpha_2 u v\cdot q}{4\pi^2 t}\Phi_{3;\pi}(\underline{\alpha})+frac{i f_{3\pi}\alpha_3 u v\cdot q}{4\pi^2 t}\Phi_{3;\pi}(\underline{\alpha})-\frac{if_{3\pi} u v\cdot q}{4\pi^2 t}\Phi_{3;\pi}(\underline{\alpha})+\frac{i f_{3\pi}\alpha_3 v\cdot q}{4\pi^2 t}\Phi_{3;\pi}(\underline{\alpha})
\\ \nonumber &&-\frac{i f_{3\pi} v\cdot q}{4\pi^2 t}\Phi_{3;\pi}(\underline{\alpha})\Big) \, ,
\\
&&G_{\Xi_c^{\prime0}[{3\over2}^-]\to\Xi_c^{*+}\pi^-}^D(\omega,\omega^{\prime})= \frac{g_{\Xi_c^{\prime0}[{3\over2}^-]\to\Xi_c^{*+}\pi^- } f_{\Xi_c^{\prime0}[{3\over2}^-]}f_{\Xi_c^{*+}}}{(\bar{\Lambda}_{\Xi_c^{\prime0}[{3\over2}^-]}-\omega^{\prime})(\bar{\Lambda}_{\Xi_c^{*+}}-\omega)}
\\ \nonumber&=& \int_0^\infty dt \int_0^1 du e^{i(1-u)\omega^\prime t} e^{iu\omega t}\times 4\times \Big(\frac{f_{\pi} m_s u}{24 \pi^2 t^2}\phi_{2;\pi}(u)+\frac{f_{\pi} m_{\pi}^2 u}{72(m_u+m_d)\pi^2 t^2}\phi_{3;\pi}^{\sigma}(u)
\\ \nonumber &&+\frac{f_{\pi} m_s u}{384 \pi^2}\phi_{4;\pi}(u)+\frac{f_{\pi} u}{72}\langle\bar s s\rangle\phi_{2;\pi}(u)+\frac{f_{\pi} m_s m_{\pi}^2 u t^2}{1728(m_u+m_d)}\langle\bar s s\rangle\phi_{3;\pi}^{\sigma}(u)+\frac{f_{\pi} u t^2}{1152}\langle\bar s s \rangle\phi_{4;\pi}(u)
\\ \nonumber &&+\frac{f_{\pi} u t^2}{1152}\langle g_s \bar s \sigma G s\rangle\phi_{2;\pi}(u)+\frac{f_{\pi} u t^4}{18432}\langle g_s \bar s \sigma G s\rangle\phi_{4;\pi}(u)\Big)
\\ \nonumber &-&\int_0^\infty dt \int_0^1 du \int \mathcal{D}\underline{\alpha}e^{i\omega^\prime t(\alpha_2+u\alpha_3)}e^{i\omega t(1-\alpha_2-u\alpha_3)}\times{1\over2}\times\Big(\frac{i f_{3\pi}\alpha_3 u^2 v\cdot q}{12\pi^2 t}\Phi_{3;\pi}(\underline{\alpha})+\frac{f_{3\pi} u v\cdot q}{12\pi^2 t}\Phi_{3;\pi} (\underline{\alpha})
\\ \nonumber &&+\frac{i f_{3\pi}\alpha_2 u v \cdot q}{12\pi^2 t}\Phi_{3;\pi}(\underline{\alpha})+\frac{i f_{3\pi}\alpha_3 u v \cdot q}{12 \pi^2}\Phi_{3;\pi}(\underline{\alpha})-\frac{i f_{3\pi} u v \cdot q}{12\pi^2 t}\Phi_{3;\pi}(\underline{\alpha})+\frac{i f_{3\pi} \alpha_2 v \cdot q}{12 \pi^2 t}\Phi_{3;\pi}(\underline{\alpha})
\\ \nonumber &&-\frac{i f_{3\pi} v \cdot q}{12\pi^2 t}\Phi_{3;\pi}(\underline{\alpha})\Big) \, ,
\\
&&G_{\Xi_c^{\prime0}[{3\over2}^-]\to\Xi_c^{*+}\pi^-}^S(\omega,\omega^{\prime})= \frac{g_{\Xi_c^{\prime0}[{3\over2}^-]\to\Xi_c^{*+}\pi^- } f_{\Xi_c^{\prime0}[{3\over2}^-]}f_{\Xi_c^{*+}}}{(\bar{\Lambda}_{\Xi_c^{\prime0}[{3\over2}^-]}-\omega^{\prime})(\bar{\Lambda}_{\Xi_c^{*+}}-\omega)}
\\ \nonumber&=& \int_0^\infty dt \int_0^1 du e^{i(1-u)\omega^\prime t} e^{iu\omega t}\times 4\times \Big(-\frac{f_{\pi} m_{\pi}^2}{6\pi^2(m_u+m_d) t^4}\phi_{3;\pi}^p(u)-\frac{i f_{\pi} m_{\pi}^2 v\cdot q}{36\pi^2(m_u+m_d) t^3}\phi_{3;\pi}^{\sigma}(u)
\\ \nonumber &&+\frac{f_{\pi} m_s}{24\pi^2 t^3 v\cdot q}\psi_{4;\pi}(u)+\frac{i f_{\pi}}{72 t v \cdot q}\langle\bar s s\rangle\psi_{4;\pi}(u)-\frac{f_{\pi}m_s m_{\pi}^2}{144(m_u+m_d)}\phi_{3;\pi}^p(u)+\frac{i f_{\pi} t}{1152 v\cdot q}\langle g_s \bar s \sigma G s\rangle\psi_{4;\pi}(u)
\\ \nonumber &&-\frac{if_{\pi} m_s m_{\pi}^2 t v\cdot q}{864(m_u+m_d)\pi^2 t^2}\phi_{3;\pi}^{\sigma}(u)\Big) \, .
\end{eqnarray}
The sum rule equations for the $\Xi_b^{\prime-}[{1\over2}^-]$ belonging to $[\mathbf{6}_F, 0 , 1, \lambda]$ are
\begin{eqnarray}
&&G_{\Xi_c^{\prime0}[{1\over2}^-]\to\Xi_c^{+}\pi^-}^S(\omega,\omega^{\prime})= \frac{g_{\Xi_c^{\prime0}[{1\over2}^-]\to\Xi_c^{+}\pi^- } f_{\Xi_c^{\prime0}[{1\over2}^-]}f_{\Xi_c^{+}}}{(\bar{\Lambda}_{\Xi_c^{\prime0}[{1\over2}^-]}-\omega^{\prime})(\bar{\Lambda}_{\Xi_c^{+}}-\omega)}
\\ \nonumber &=& \int_0^\infty dt \int_0^1 du e^{i(1-u)\omega^\prime t} e^{iu\omega t}\times 4\times \Big(-\frac{3 f_{\pi} m_{\pi}^2}{4\pi^2(m_u+m_d) t^4}\phi_{3;\pi}^p(u)-\frac{i f_{\pi} m_{\pi}^2 v \cdot q}{8\pi^2(m_u+m_d) t^3}\phi_{3;\pi}^{\sigma}(u)
\\ \nonumber &&+\frac{i f_{\pi}}{16 t v \cdot q}\langle \bar s s \rangle\psi_{4;\pi}(u)+\frac{i f_{\pi} t}{256 v \cdot q}\langle g_s \bar s \sigma G s\rangle\psi_{4;\pi}(u)+\frac{3 if_{\pi} m_s}{16\pi^2 t^3 v \cdot q}\psi_{4;\pi}(u)-\frac{f_{\pi} m_s m_{\pi}^2}{32 (m_u+m_d)}\langle \bar s s\rangle\phi_{3;\pi}^p(u)
\\ \nonumber &&-\frac{if_{\pi} m_s m_{\pi}^2 t v \cdot q}{192(m_u+m_d)}\langle \bar s s\rangle\phi_{3;\pi}^{\sigma}(u)\Big) \, ,
\\
&&G_{\Xi_c^{\prime0}[{1\over2}^-]\to\Lambda_c^{+}K^-}^S(\omega,\omega^{\prime})= \frac{g_{\Xi_c^{\prime0}[{1\over2}^-]\to\Lambda_c^{+}K^- } f_{\Xi_c^{\prime0}[{1\over2}^-]}f_{\Lambda_c^{+}}}{(\bar{\Lambda}_{\Xi_c^{\prime0}[{1\over2}^-]}-\omega^{\prime})(\bar{\Lambda}_{\Lambda_c^{+}}-\omega)}
\\ \nonumber &=& \int_0^\infty dt \int_0^1 du e^{i(1-u)\omega^\prime t} e^{iu\omega t}\times 4\times \Big(-\frac{3 f_{K} m_{K}^2}{4\pi^2(m_u+m_s) t^4}\phi_{3;K}^p(u)-\frac{i f_{K} m_{K}^2 v \cdot q}{8\pi^2(m_u+m_s) t^3}\phi_{3;K}^{\sigma}(u)
\\ \nonumber &&+\frac{i f_{K}}{16 t v \cdot q}\langle \bar q q \rangle\psi_{4;K}(u)+\frac{i f_{K} t}{256 v \cdot q}\langle g_s \bar q \sigma G q\rangle\psi_{4;K}(u)\Big) \, .
\end{eqnarray}
The sum rule equations for the $\Xi_b^{\prime-}[{1\over2}^-]$ belonging to $[\mathbf{6}_F, 1 , 1, \lambda]$ are
\begin{eqnarray}
&&G_{\Xi_c^{\prime0}[{1\over2}^-]\to\Xi_c^{*+}\pi^-}^D(\omega,\omega^{\prime})= \frac{g_{\Xi_c^{\prime0}[{1\over2}^-]\to\Xi_c^{*+}\pi^- } f_{\Xi_c^{\prime0}[{1\over2}^-]}f_{\Xi_c^{*+}}}{(\bar{\Lambda}_{\Xi_c^{\prime0}[{1\over2}^-]}-\omega^{\prime})(\bar{\Lambda}_{\Xi_c^{*+}}-\omega)}
\\ \nonumber&=& \int_0^\infty dt \int_0^1 du e^{i(1-u)\omega^\prime t} e^{iu\omega t}\times 4\times \Big(-\frac{f_{\pi} u}{4\pi^2 t^3}\phi_{2;\pi}(u)+\frac{f_{\pi} m_s m_{\pi}^2 u}{48(m_u+m_d) \pi^2 t}\phi_{3;\pi}^{\sigma}(u)
\\ \nonumber &&-\frac{f_{\pi} u}{64 \pi^2 t}\phi_{4;\pi}(u)-\frac{f_{\pi} m_s u t}{96}\langle\bar s s\rangle \phi_{2;\pi}(u)+\frac{f_{\pi} m_{\pi}^2 u t}{144(m_u+m_d)}\langle \bar s s \rangle \phi_{3;\pi}^{\sigma}(u)-\frac{f_{\pi} m_s t^3}{1536}\langle\bar s s \rangle \phi_{4;\pi}(u)
\\ \nonumber &&+\frac{f_{\pi} m_{\pi}^2 t^3}{2304(m_u+m_d)}\langle g_s \bar s \sigma G s \rangle\phi_{3;\pi}^{\sigma}(u)\Big)
\\ \nonumber &-&\int_0^\infty dt \int_0^1 du \int \mathcal{D}\underline{\alpha}e^{i\omega^\prime t(\alpha_2+u\alpha_3)}e^{i\omega t(1-\alpha_2-u\alpha_3)}\times{1\over2}\times\Big(-\frac{f_{\pi}\alpha_3 u^2}{8\pi^2 t}\Phi_{4;\pi}(\underline{\alpha})-\frac{f_{\pi}\alpha_2 u}{8\pi^2 t}\Phi_{4;\pi}(\underline{\alpha})
\\ \nonumber &&-\frac{f_{\pi}\alpha_3 u}{16\pi^2 t}\Phi_{4;\pi}(\underline{\alpha})-\frac{f_{\pi}\alpha_3 u}{16\pi^2 t}\widetilde\Phi_{4;\pi}(\underline{\alpha})+\frac{f_{\pi} u}{8\pi^2 t}\Phi_{4;\pi}(\underline{\alpha})-\frac{f_{\pi}\alpha_2}{16\pi^2 t}\Phi_{4;\pi}(\underline{\alpha})
-\frac{f_{\pi}\alpha_2}{16\pi^2 t}\widetilde\Phi{4;\pi}(\underline{\alpha})
\\ \nonumber &&+\frac{f_{\pi}}{16\pi^2 t}\Phi_{4;\pi}(\underline{\alpha})+\frac{f_{\pi}}{16\pi^2 t}\widetilde\Phi_{4;\pi}(\underline{\alpha})+\frac{if_{\pi} u}{8\pi^2 t^2 v\cdot q}\Psi_{4;\pi}(\underline{\alpha})
+\frac{3 i f_{\pi}}{8\pi^2 t^2 v\cdot q}\widetilde\Psi_{4;\pi}(\underline{\alpha})
-\frac{i f_{\pi}}{8\pi^2  t^2 v \cdot q}\Phi_{4;\pi}(\underline{\alpha})
\\ \nonumber &&-\frac{3 i f_{\pi}}{8\pi^2 t^2 v\cdot q}\widetilde\Phi_{4;\pi}(\underline{\alpha})+\frac{i f_{\pi}}{8\pi^2 t^2 v\cdot q}\Psi_{4;\pi}(\underline{\alpha})
-\frac{i f_{\pi}}{8\pi^2 t^2 v\cdot q}\widetilde\Psi_{4;\pi}(\underline{\alpha})\Big) \, ,
\\
&&G_{\Xi_c^{\prime0}[{1\over2}^-]\to\Xi_c^{\prime+}\pi^-}^S(\omega,\omega^{\prime})= \frac{g_{\Xi_c^{\prime0}[{1\over2}^-]\to\Xi_c^{\prime+}\pi^- } f_{\Xi_c^{\prime0}[{1\over2}^-]}f_{\Xi_c^{\prime+}}}{(\bar{\Lambda}_{\Xi_c^{\prime0}[{1\over2}^-]}-\omega^{\prime})(\bar{\Lambda}_{\Xi_c^{\prime+}}-\omega)}
\\ \nonumber&=& \int_0^\infty dt \int_0^1 du e^{i(1-u)\omega^\prime t} e^{iu\omega t}\times 4\times \Big(\frac{3 i f_{\pi} v\cdot q}{2\pi^2 t^4}\phi_{2;\pi}(u)+\frac{3i f_{\pi} v\cdot q}{32\pi^2 t^2}\phi_{4;\pi}(u)-\frac{i f_{\pi} m_{\pi}^2 v\cdot q}{24(m_u+m_d)}\phi_{3;\pi}^{\sigma}(u)
\\ \nonumber &&-\frac{i f_{\pi} m_{\pi}^2 v\cdot q t^2}{384(m_u+m_d)}\langle g_s \bar s\sigma G s\rangle\phi_{3;\pi}^{\sigma}(u)+\frac{if_{\pi}m_s v\cdot q}{16}\langle\bar s s\rangle\phi_{2;\pi}(u)-\frac{if_{\pi}m_s m_{\pi}^2 v\cdot q}{8\pi^2(m_u+m_d)t^2}\phi_{3;\pi}^{\sigma}(u)
\\ \nonumber &&+\frac{if_{\pi}m_s t^2 v\cdot q}{256}\langle\bar s s\rangle\phi_{4;\pi}(u)\Big)
\\ \nonumber &-&\int_0^\infty dt \int_0^1 du \int \mathcal{D}\underline{\alpha}e^{i\omega^\prime t(\alpha_2+u\alpha_3)}e^{i\omega t(1-\alpha_2-u\alpha_3)}\times{1\over2}\times\Big(-\frac{3if_{\pi} u v\cdot q}{4\pi^2 t^2}\Phi_{4;\pi}(\underline{\alpha})+\frac{i f_{\pi} u v\cdot q}{2\pi^2 t^2}\Psi_{4;\pi}(\underline{\alpha})
\\ \nonumber &&-\frac{i f_{\pi} v\cdot q}{8\pi^2 t^2}\Phi_{4;\pi}(\underline{\alpha})-\frac{3if_{\pi} v\cdot q}{8\pi^2 t^2}\widetilde\Phi_{4;\pi}(\underline{\alpha})-\frac{i f_{\pi} v\cdot q}{4\pi^2 t^2}\Psi_{4;\pi}(\underline{\alpha})+\frac{i f_{\pi} v\cdot q}{4\pi^2 t^2}\widetilde\Psi_{4;\pi}(\underline{\alpha})\Big) \, .
\end{eqnarray}
The sum rule equations for the $\Xi_b^{\prime-}[{3\over2}^-]$ belonging to $[\mathbf{6}_F, 1 , 1, \lambda]$ are
\begin{eqnarray}
&&G_{\Xi_c^{\prime0}[{3\over2}^-]\to\Xi_c^{\prime+}\pi^-}^D(\omega,\omega^{\prime})= \frac{g_{\Xi_c^{\prime0}[{3\over2}^-]\to\Xi_c^{\prime+}\pi^- } f_{\Xi_c^{\prime0}[{3\over2}^-]}f_{\Xi_c^{\prime+}}}{(\bar{\Lambda}_{\Xi_c^{\prime0}[{3\over2}^-]}-\omega^{\prime})(\bar{\Lambda}_{\Xi_c^{\prime+}}-\omega)}
\\ \nonumber &=& \int_0^\infty dt \int_0^1 du e^{i(1-u)\omega^\prime t} e^{iu\omega t}\times 4\times \Big(\frac{i f_{\pi} u}{4\pi^2 t^3}\phi_{2;\pi}(u)-\frac{if_{\pi} m_s m_{\pi}^2 u}{48 (m_u+m_d)\pi^2 t}\phi_{3;\pi}^{\sigma}(u)
\\ \nonumber &&+\frac{i f_{\pi} u}{64\pi^2 t}\phi_{4;\pi}(u)+\frac{i f_{\pi} m_s u t}{96}\langle\bar s s\rangle\phi_{2;\pi}(u)-\frac{i f_{\pi} m_{\pi}^2 u t}{144(m_u+m_d)}\langle\bar s s\rangle\phi_{3;\pi}^{\sigma}(u)+\frac{i f_{\pi} m_s u t^3}{1536}\langle\bar s s\rangle\phi_{4;\pi}(u)
\\ \nonumber &&-\frac{i f_{\pi} m_{\pi}^2 u t^3}{2304(m_u+m_d)}\langle g_s\bar s \sigma G s\rangle\phi_{3;\pi}^{\sigma}(u)\Big)
\\ \nonumber &-&\int_0^\infty dt \int_0^1 du \int \mathcal{D}\underline{\alpha}e^{i\omega^\prime t(\alpha_2+u\alpha_3)}e^{i\omega t(1-\alpha_2-u\alpha_3)}\times{1\over2}\times\Big(\frac{i f_{\pi}\alpha_3 u^2}{8\pi^2 t}\Phi_{4;\pi}(\underline{\alpha})+\frac{i f_{\pi}\alpha_2 u}{8\pi^2 t}\Phi_{4;\pi}(\underline{\alpha})
\\ \nonumber &&+\frac{i f_{\pi}\alpha_3 u}{16\pi^2 t}\Phi_{4;\pi}(\underline{\alpha})+\frac{i f_{\pi}\alpha_3 u}{16\pi^2 t}\widetilde\Phi_{4;\pi}(\underline{\alpha})-\frac{i f_{\pi} u}{8\pi^2 t}\Phi_{4;\pi}(\underline{\alpha})+\frac{i f_{\pi}\alpha_2}{16\pi^2 t}\Phi_{4;\pi}(\underline{\alpha})+\frac{i f_{\pi}\alpha_2}{16\pi^2 t}\widetilde\Phi_{4;\pi}(\underline{\alpha})
\\ \nonumber &&-\frac{i f_{\pi}}{16\pi^2 t}\Phi_{4;\pi}(\underline{\alpha})-\frac{i f_{\pi}}{16\pi^2 t}\widetilde\Phi_{4;\pi}(\underline{\alpha})+\frac{f_{\pi} u}{8\pi^2 t^2 v\cdot q}\Psi_{4;\pi}(\underline{\alpha})+\frac{3 f_{\pi} u}{8\pi^2 t^2 v\cdot q}\widetilde\Psi_{4;\pi}(\underline{\alpha})-\frac{f_{\pi}}{8\pi^2 t^2 v\cdot q}\Phi_{4;\pi}(\underline{\alpha})
\\ \nonumber &&-\frac{3 f_{\pi}}{8\pi^2 t^2 v\cdot q}\widetilde\Phi_{4;\pi}(\underline{\alpha})+\frac{f_{\pi}}{8\pi^2 t^2 v\cdot q}\Psi_{4;\pi}(\underline{\alpha})-\frac{f_{\pi}}{8\pi^2 t^2 v\cdot q}\widetilde\Psi_{4;\pi}(\underline{\alpha})\Big) \, ,
\\
&&G_{\Xi_c^{\prime0}[{3\over2}^-]\to\Sigma_c^{+}K^-}^D(\omega,\omega^{\prime})= \frac{g_{\Xi_c^{\prime0}[{3\over2}^-]\to\Sigma_c^{+}K^- } f_{\Xi_c^{\prime0}[{3\over2}^-]}f_{\Sigma_c^{+}}}{(\bar{\Lambda}_{\Xi_c^{\prime0}[{3\over2}^-]}-\omega^{\prime})(\bar{\Lambda}_{\Sigma_c^{+}}-\omega)}
\\ \nonumber &=& \int_0^\infty dt \int_0^1 du e^{i(1-u)\omega^\prime t} e^{iu\omega t}\times 4\times \Big(\frac{i f_{K} u}{4\pi^2 t^3}\phi_{2;K}(u)+\frac{i f_{K} u}{64\pi^2 t}\phi_{4;K}(u)-\frac{i f_{K} m_{K}^2 u t}{144(m_u+m_s)}\langle\bar q q\rangle\phi_{3;K}^{\sigma}(u)
\\ \nonumber &&-\frac{i f_{K} m_{K}^2 u t^3}{2304(m_u+m_s)}\langle g_s\bar q \sigma G q\rangle\phi_{3;K}^{\sigma}(u)\Big)
\\ \nonumber &-&\int_0^\infty dt \int_0^1 du \int \mathcal{D}\underline{\alpha}e^{i\omega^\prime t(\alpha_2+u\alpha_3)}e^{i\omega t(1-\alpha_2-u\alpha_3)}\times{1\over2}\times\Big(\frac{i f_{K}\alpha_3 u^2}{8\pi^2 t}\Phi_{4;K}(\underline{\alpha})+\frac{i f_{K}\alpha_2 u}{8\pi^2 t}\Phi_{4;K}(\underline{\alpha})
\\ \nonumber &&+\frac{i f_{K}\alpha_3 u}{16\pi^2 t}\Phi_{4;K}(\underline{\alpha})+\frac{i f_{K}\alpha_3 u}{16\pi^2 t}\widetilde\Phi_{4;K}(\underline{\alpha})-\frac{i f_{K} u}{8\pi^2 t}\Phi_{4;K}(\underline{\alpha})+\frac{i f_{K}\alpha_2}{16\pi^2 t}\Phi_{4;K}(\underline{\alpha})+\frac{i f_{K}\alpha_2}{16\pi^2 t}\widetilde\Phi_{4;K}(\underline{\alpha})
\\ \nonumber &&-\frac{i f_{K}}{16\pi^2 t}\Phi_{4;K}(\underline{\alpha})-\frac{i f_{K}}{16\pi^2 t}\widetilde\Phi_{4;\pi}(\underline{\alpha})+\frac{f_{\pi} u}{8\pi^2 t^2 v\cdot q}\Psi_{4;K}(\underline{\alpha})+\frac{3 f_{K} u}{8\pi^2 t^2 v\cdot q}\widetilde\Psi_{4;K}(\underline{\alpha})-\frac{f_{K}}{8\pi^2 t^2 v\cdot q}\Phi_{4;K}(\underline{\alpha})
\\ \nonumber &&-\frac{3 f_{K}}{8\pi^2 t^2 v\cdot q}\widetilde\Phi_{4;K}(\underline{\alpha})+\frac{f_{K}}{8\pi^2 t^2 v\cdot q}\Psi_{4;K}(\underline{\alpha})-\frac{f_{K}}{8\pi^2 t^2 v\cdot q}\widetilde\Psi_{4;K}(\underline{\alpha})\Big) \, ,
\\
&&G_{\Xi_c^{\prime0}[{3\over2}^-]\to\Xi_c^{*+}\pi^-}^S(\omega,\omega^{\prime})= \frac{g_{\Xi_c^{\prime0}[{3\over2}^-]\to\Xi_c^{*+}\pi^- } f_{\Xi_c^{\prime0}[{3\over2}^-]}f_{\Xi_c^{*+}}}{(\bar{\Lambda}_{\Xi_c^{\prime0}[{3\over2}^-]}-\omega^{\prime})(\bar{\Lambda}_{\Xi_c^{*+}}-\omega)}
\\ \nonumber&=& \int_0^\infty dt \int_0^1 du e^{i(1-u)\omega^\prime t} e^{iu\omega t}\times 4\times \Big(\frac{f_{\pi} v\cdot q}{3\pi^2 t^4}\phi_{2;\pi}(u)-\frac{f_{\pi} v\cdot q}{48\pi^2 t^2}\phi_{4;\pi}(u)
\\ \nonumber &&+\frac{f_{\pi} m_{\pi}^2 v\cdot q}{108(m_u+m_d)}\langle\bar s s\rangle\phi_{3;\pi}^{\sigma}(u)+\frac{f_{\pi} m_{\pi}^2 t^2 v\cdot q}{1728(m_u+m_d)}\langle g_s \bar s \sigma G s\rangle\phi_{3;\pi}^{\sigma}(u)-\frac{f_{\pi} m_s v\cdot q}{72}\langle \bar s s\rangle\phi_{2;\pi}(u)
\\ \nonumber &&-\frac{f_{\pi} m_s t^2 v\cdot q}{1152}\langle\bar s s\rangle\phi_{4;\pi}(u)+\frac{f_{\pi} m_s m_{\pi}^2 v\cdot q}{36\pi^2(m_u+m_d) t^2}\phi_{3;\pi}^{\sigma}(u)\Big)
\\ \nonumber &-&\int_0^\infty dt \int_0^1 du \int \mathcal{D}\underline{\alpha}e^{i\omega^\prime t(\alpha_2+u\alpha_3)}e^{i\omega t(1-\alpha_2-u\alpha_3)}\times{1\over2}\times\Big(\frac{f_{\pi} u v\cdot q}{6\pi^2 t^2}\Phi_{4;\pi}(\underline{\alpha})-\frac{7f_{\pi} u v\cdot q}{72\pi^2 t^2}\Psi_{4;\pi}(\underline{\alpha})
\\ \nonumber &&+\frac{f_{\pi} u v\cdot q}{24\pi^2 t^2}\widetilde\Psi_{4;\pi}(\underline{\alpha})+\frac{f_{\pi} v\cdot q}{72\pi^2 t^2}\Phi_{4;\pi}(\underline{\alpha})+\frac{f_{\pi} v\cdot q}{24 \pi^2 t^2}\widetilde\Phi_{4;\pi}(\underline{\alpha})+\frac{5 f_{\pi} v\cdot q}{72\pi^2 t^2}\Psi_{4;\pi}(\underline{\alpha})-\frac{5 f_{\pi} v\cdot q}{72\pi^2 t^2}\widetilde\Psi_{4;\pi}(\underline{\alpha})\Big) \, .
\end{eqnarray}
The sum rule equations for the $\Xi_b^{\prime-}[{3\over2}^-]$ belonging to $[\mathbf{6}_F, 2 , 1, \lambda]$ are
\begin{eqnarray}
&&G_{\Xi_c^{\prime0}[{3\over2}^-]\to\Xi_c^{+}\pi^-}^D(\omega,\omega^{\prime})= \frac{g_{\Xi_c^{\prime0}[{3\over2}^-]\to\Xi_c^{+}\pi^- } f_{\Xi_c^{\prime0}[{3\over2}^-]}f_{\Xi_c^{+}}}{(\bar{\Lambda}_{\Xi_c^{\prime0}[{3\over2}^-]}-\omega^{\prime})(\bar{\Lambda}_{\Xi_c^{+}}-\omega)}
\\ \nonumber &=& \int_0^\infty dt \int_0^1 du e^{i(1-u)\omega^\prime t} e^{iu\omega t}\times 4\times \Big(\frac{f_{\pi} m_s u}{4\pi^2 t^2}\phi_{2;\pi}(u)+\frac{f_{\pi} m_{\pi}^2 u}{12(m_u+m_d)\pi^2 t^2}\phi_{3;\pi}^{\sigma}(u)
\\ \nonumber &&+\frac{f_{\pi} m_s u}{64\pi^2}\phi_{4;\pi}(u)+\frac{f_{\pi} u}{12}\langle\bar s s\rangle\phi_{2;\pi}(u)+\frac{f_{\pi} m_s u t^2}{288(m_u+m_d)}\langle\bar s s\rangle\phi_{3;\pi}^{\sigma}(u)+\frac{f_{\pi} u t^2}{192}\langle\bar s s\rangle\phi_{4;\pi}(u)
\\ \nonumber &&+\frac{f_{\pi} u t^2}{192}\langle g_s \bar s \sigma G s\rangle\phi_{2;\pi}(u)+\frac{f_{\pi} u t^4}{3072}\langle g_s \bar s \sigma G s\rangle\phi_{4;\pi}(u)\Big)
\\ \nonumber  &-&\int_0^\infty dt \int_0^1 du \int \mathcal{D}\underline{\alpha}e^{i\omega^\prime t(\alpha_2+u\alpha_3)}e^{i\omega t(1-\alpha_2-u\alpha_3)}\times{1\over2}\times\Big(\frac{f_{3\pi} u}{2\pi t^2}\Phi_{3;\pi}(\underline{\alpha})-\frac{f_{3\pi}}{2\pi^2 t^2}\Phi_{3;\pi}(\underline{\alpha})
\\ \nonumber &&+\frac{i f_{3\pi} u^2 \alpha_3 v\cdot q}{2\pi^2 t}\Phi_{3;\pi}(\underline{\alpha})+\frac{i f_{3\pi} u \alpha_2 v\cdot q}{2\pi^2 t}\Phi_{3;\pi}(\underline{\alpha})-\frac{i f_{3\pi} u v\cdot q}{2\pi^2 t}\Phi_{3;\pi}(\underline{\alpha})\Big) \, ,
\\
&&G_{\Xi_c^{\prime0}[{3\over2}^-]\to\Lambda_c^{+}K^-}^D(\omega,\omega^{\prime})= \frac{g_{\Xi_c^{\prime0}[{3\over2}^-]\to\Lambda_c^{+}K^- } f_{\Xi_c^{\prime0}[{3\over2}^-]}f_{\Lambda_c^{+}}}{(\bar{\Lambda}_{\Xi_c^{\prime0}[{3\over2}^-]}-\omega^{\prime})(\bar{\Lambda}_{\Lambda_c^{+}}-\omega)}
\\ \nonumber &=& \int_0^\infty dt \int_0^1 du e^{i(1-u)\omega^\prime t} e^{iu\omega t}\times 4\times \Big(\frac{f_{K} m_{K}^2 u}{12(m_u+m_s)\pi^2 t^2}\phi_{3;K}^{\sigma}(u)+\frac{f_{K} u}{12}\langle\bar q q\rangle\phi_{2;K}(u)
\\ \nonumber &&+\frac{f_{K} u t^2}{192}\langle\bar q q\rangle\phi_{4;K}(u)+\frac{f_{K} u t^2}{192}\langle g_s \bar q \sigma G q\rangle\phi_{2;K}(u)+\frac{f_{K} u t^4}{3072}\langle g_s \bar q \sigma G q\rangle\phi_{4;K}(u)\Big)
\\ \nonumber  &-&\int_0^\infty dt \int_0^1 du \int \mathcal{D}\underline{\alpha}e^{i\omega^\prime t(\alpha_2+u\alpha_3)}e^{i\omega t(1-\alpha_2-u\alpha_3)}\times{1\over2}\times\Big(\frac{f_{3 K} u}{2\pi t^2}\Phi_{3;K}(\underline{\alpha})-\frac{f_{3 K}}{2\pi^2 t^2}\Phi_{3;K}(\underline{\alpha})
\\ \nonumber &&+\frac{i f_{3 K} u^2 \alpha_3 v\cdot q}{2\pi^2 t}\Phi_{3;K}(\underline{\alpha})+\frac{i f_{3 K} u \alpha_2 v\cdot q}{2\pi^2 t}\Phi_{3;K}(\underline{\alpha})-\frac{i f_{3 K} u v\cdot q}{2\pi^2 t}\Phi_{3;K}(\underline{\alpha})\Big) \, ,
\\
&&G_{\Xi_c^{\prime0}[{3\over2}^-]\to\Xi_c^{\prime+}\pi^-}^D(\omega,\omega^{\prime})= \frac{g_{\Xi_c^{\prime0}[{3\over2}^-]\to\Xi_c^{\prime+}\pi^- } f_{\Xi_c^{\prime0}[{3\over2}^-]}f_{\Xi_c^{\prime+}}}{(\bar{\Lambda}_{\Xi_c^{\prime0}[{3\over2}^-]}-\omega^{\prime})(\bar{\Lambda}_{\Xi_c^{\prime+}}-\omega)}
\\ \nonumber &=& \int_0^\infty dt \int_0^1 du e^{i(1-u)\omega^\prime t} e^{iu\omega t}\times 4\times \Big(\frac{3i f_{\pi} u}{4\pi^2 t^3}\phi_{2;\pi}(u)-\frac{i f_{\pi} m_{\pi}^2 m_s u}{16 (m_u+m_d)\pi^2 t}\phi_{3;\pi}^{\sigma}(u)
\\ \nonumber &&+\frac{3 i f_{\pi} u}{64\pi^2 t}\phi_{4;\pi}(u)+\frac{i f_{\pi} m_s u t}{32}\langle\bar s s\rangle\phi_{2;\pi}(u)-\frac{i f_{\pi} m_{\pi}^2 u t}{48(m_u+m_d)}\langle\bar s s\rangle\phi_{3;\pi}^{\sigma}(u)+\frac{i f_{\pi} m_s u t^3}{512}\langle\bar s s\rangle\phi_{4;\pi}(u)
\\ \nonumber &&-\frac{i f_{\pi} m_{\pi}^2 u t^3}{768(m_u+m_d)}\langle g_s \bar s \sigma G s\rangle\phi_{3;\pi}^{sigma}(u)\Big)
\\ \nonumber &-&\int_0^\infty dt \int_0^1 du \int \mathcal{D}\underline{\alpha}e^{i\omega^\prime t(\alpha_2+u\alpha_3)}e^{i\omega t(1-\alpha_2-u\alpha_3)}\times{1\over2}\times\Big(\frac{3f_{\pi} u}{8\pi^2 t^2 v\cdot q}\Psi_{4;\pi}(\underline{\alpha})-\frac{3 f_{\pi} u}{8\pi^2 t62 v\cdot q}\widetilde\Psi_{4;\pi}(\underline{\alpha})
\\ \nonumber &&-\frac{3 f_{\pi}}{8\pi^2 t^2 v\cdot q}\Phi_{4;\pi}(\underline{\alpha})+\frac{3 f_{\pi}}{8\pi^2 t^2 v\cdot q}\widetilde\Phi_{4;\pi}(\underline{\alpha})-\frac{3 f_{\pi} }{8\pi^2 t^2 v\cdot q}\Psi_{4;\pi}(\underline{\alpha})+\frac{3 f_{\pi}}{8\pi^2 t^2 v\cdot q}\widetilde\Psi_{4;\pi}(\underline{\alpha})
\\ \nonumber &&+\frac{3 i f_{\pi}\alpha_3 u^2}{8\pi^2 t}\Phi_{4;\pi}(\underline{\alpha})+\frac{3 i f_{\pi}\alpha_2 u}{8\pi^2 t}\Phi_{4;\pi}(\underline{\alpha})+\frac{3 i f_{\pi}\alpha_3 u}{16\pi^2 t}\Phi_{4;\pi}(\underline{\alpha})+\frac{3 i f_{\pi}\alpha_3 u}{16\pi^2 t}\widetilde\Phi_{4;\pi}(\underline{\alpha})
\\ \nonumber &&-\frac{3 i f_{\pi} u}{8\pi^2 t}\Phi_{4;\pi}(\underline{\alpha})+\frac{3 i f_{\pi}\alpha_2}{16\pi^2 t}\Phi_{4;\pi}(\underline{\alpha})+\frac{3 if_{\pi}\alpha_2}{16\pi^2 t}\widetilde\Phi_{4;\pi}(\underline{\alpha})-\frac{3 i f_{\pi}}{16\pi^2 t}\Phi_{4;\pi}(\underline{\alpha})-\frac{3 i f_{\pi}}{16\pi^2 t}\widetilde\Phi_{4;\pi}(\underline{\alpha})\Big) \, ,
\\
&&G_{\Xi_c^{\prime0}[{3\over2}^-]\to\Sigma_c^{+}K^-}^D(\omega,\omega^{\prime})= \frac{g_{\Xi_c^{\prime0}[{3\over2}^-]\to\Sigma_c^{+}K^- } f_{\Xi_c^{\prime0}[{3\over2}^-]}f_{\Sigma_c^{+}}}{(\bar{\Lambda}_{\Xi_c^{\prime0}[{3\over2}^-]}-\omega^{\prime})(\bar{\Lambda}_{\Sigma_c^{+}}-\omega)}
\\ \nonumber &=& \int_0^\infty dt \int_0^1 du e^{i(1-u)\omega^\prime t} e^{iu\omega t}\times 4\times \Big(\frac{3i f_{K} u}{4\pi^2 t^3}\phi_{2;K}(u)+\frac{3 i f_{K} u}{64\pi^2 t}\phi_{4;K}(u)-\frac{i f_{K} m_{K}^2 u t}{48(m_u+m_s)}\langle\bar q q\rangle\phi_{3;K}^{\sigma}(u)
\\ \nonumber &&-\frac{i f_{K} m_{K}^2 u t^3}{768(m_u+m_s)}\langle g_s \bar q \sigma G q\rangle\phi_{3;K}^{sigma}(u)\Big)
\\ \nonumber &-&\int_0^\infty dt \int_0^1 du \int \mathcal{D}\underline{\alpha}e^{i\omega^\prime t(\alpha_2+u\alpha_3)}e^{i\omega t(1-\alpha_2-u\alpha_3)}\times{1\over2}\times\Big(\frac{3f_{K} u}{8\pi^2 t^2 v\cdot q}\Psi_{4;K}(\underline{\alpha})-\frac{3 f_{K} u}{8\pi^2 t62 v\cdot q}\widetilde\Psi_{4;K}(\underline{\alpha})
\\ \nonumber &&-\frac{3 f_{K}}{8\pi^2 t^2 v\cdot q}\Phi_{4;K}(\underline{\alpha})+\frac{3 f_{K}}{8\pi^2 t^2 v\cdot q}\widetilde\Phi_{4;K}(\underline{\alpha})-\frac{3 f_{K} }{8\pi^2 t^2 v\cdot q}\Psi_{4;K}(\underline{\alpha})+\frac{3 f_{K}}{8\pi^2 t^2 v\cdot q}\widetilde\Psi_{4;K}(\underline{\alpha})
\\ \nonumber &&+\frac{3 i f_{K}\alpha_3 u^2}{8\pi^2 t}\Phi_{4;K}(\underline{\alpha})+frac{3 i f_{K}\alpha_2 u}{8\pi^2 t}\Phi_{4;K}(\underline{\alpha})+\frac{3 i f_{K}\alpha_3 u}{16\pi^2 t}\Phi_{4;K}(\underline{\alpha})+\frac{3 i f_{K}\alpha_3 u}{16\pi^2 t}\widetilde\Phi_{4;K}(\underline{\alpha})
\\ \nonumber &&-\frac{3 i f_{K} u}{8\pi^2 t}\Phi_{4;K}(\underline{\alpha})+\frac{3 i f_{K}\alpha_2}{16\pi^2 t}\Phi_{4;K}(\underline{\alpha})+\frac{3 if_{K}\alpha_2}{16\pi^2 t}\widetilde\Phi_{4;K}(\underline{\alpha})-\frac{3 i f_{K}}{16\pi^2 t}\Phi_{4;K}(\underline{\alpha})-\frac{3 i f_{K}}{16\pi^2 t}\widetilde\Phi_{4;K}(\underline{\alpha})\Big) \, ,
\\
&&G_{\Xi_c^{\prime0}[{3\over2}^-]\to\Xi_c^{*+}\pi^-}^D(\omega,\omega^{\prime})= \frac{g_{\Xi_c^{\prime0}[{3\over2}^-]\to\Xi_c^{*+}\pi^- } f_{\Xi_c^{\prime0}[{3\over2}^-]}f_{\Xi_c^{*+}}}{(\bar{\Lambda}_{\Xi_c^{\prime0}[{3\over2}^-]}-\omega^{\prime})(\bar{\Lambda}_{\Xi_c^{*+}}-\omega)}
\\ \nonumber&=& \int_0^\infty dt \int_0^1 du e^{i(1-u)\omega^\prime t} e^{iu\omega t}\times 4\times \Big(-\frac{i f_{\pi} u}{4\pi^2 t^3}\phi_{2;\pi}(u)+\frac{i f_{\pi} m_s m_{\pi}^2 u}{48(m_u+m_d)\pi^2 t}\phi_{3;\pi}^{\sigma}(u)
\\ \nonumber &&-\frac{i f_{\pi} u}{64\pi^2 t}\phi_{4;\pi}(u)-\frac{i f_{\pi} m_s u t}{96}\langle\bar s s\rangle\phi_{2;\pi}(u)+\frac{i f_{\pi} m_{\pi}^2 u t}{144(m_u+m_d)}\langle\bar s s\rangle\phi_{3;\pi}^{\sigma}(u)-\frac{i f_{\pi} m_s u t^3}{1536}\langle\bar s s\rangle\phi_{4;\pi}(u)
\\ \nonumber &&+\frac{i f_{\pi} m_{\pi}^2 u t^3}{2304(m_u+m_d)}\langle g_s \bar s \sigma G s\rangle\phi_{3;\pi}^{\sigma}(u)\Big)
\\ \nonumber &-&\int_0^\infty dt \int_0^1 du \int \mathcal{D}\underline{\alpha}e^{i\omega^\prime t(\alpha_2+u\alpha_3)}e^{i\omega t(1-\alpha_2-u\alpha_3)}\times{1\over2}\times\Big(-\frac{f_{\pi} u}{8\pi^2 t^2 v\cdot q}\Psi_{4;\pi}(\underline{\alpha})+\frac{f_{\pi} u}{8\pi^2 t^2 v\cdot q}\widetilde\Psi_{4;\pi}(\underline{\alpha})
\\ \nonumber &&+\frac{f_{\pi}}{8\pi^2 t^2 v\cdot q}\Phi_{4;\pi}(\underline{\alpha})-\frac{f_{\pi}}{8\pi^2 t^2 v\cdot q}\widetilde\Phi_{4;\pi}(\underline{\alpha})+\frac{f_{\pi}}{8\pi^2 t^2 v\cdot q}\Psi_{4;\pi}(\underline{\alpha})-\frac{f_{\pi}}{8\pi^2 t^2 v\cdot q}\widetilde\Psi_{4;\pi}(\underline{\alpha})
\\ \nonumber &&-\frac{i f_{\pi}\alpha_3 u^2}{8\pi^2 t}\Phi_{4;\pi}(\underline{\alpha})-\frac{i f_{\pi}\alpha_2 u}{8\pi^2 t}\Phi_{4;\pi}(\underline{\alpha})-\frac{i f_{\pi}\alpha_3 u}{16\pi^2 t}\Phi_{4;\pi}(\underline{\alpha})-\frac{i f_{\pi}\alpha_3 u}{16\pi^2 t}\widetilde\Phi_{4;\pi}(\underline{\alpha})
\\ \nonumber &&+\frac{i f_{\pi} u}{8\pi^2 t}\Phi_{4;\pi}(\underline{\alpha})-\frac{i f_{\pi}\alpha_2}{16\pi^2 t}\Phi_{4;\pi}(\underline{\alpha})-\frac{i f_{\pi}\alpha_2}{16\pi^2 t}\widetilde\Phi_{4;\pi}(\underline{\alpha})+\frac{i f_{\pi}}{16\pi^2 t}\Phi_{4;\pi}(\underline{\alpha})+\frac{i f_{\pi}}{16\pi^2 t}\widetilde\Phi_{4;\pi}(\underline{\alpha})\Big) \, ,
\\
&&G_{\Xi_c^{\prime0}[{3\over2}^-]\to\Xi_c^{*+}\pi^-}^S(\omega,\omega^{\prime})= \frac{g_{\Xi_c^{\prime0}[{3\over2}^-]\to\Xi_c^{*+}\pi^- } f_{\Xi_c^{\prime0}[{3\over2}^-]}f_{\Xi_c^{*+}}}{(\bar{\Lambda}_{\Xi_c^{\prime0}[{3\over2}^-]}-\omega^{\prime})(\bar{\Lambda}_{\Xi_c^{*+}}-\omega)}
\\ \nonumber &=& -\int_0^\infty dt \int_0^1 du \int \mathcal{D}\underline{\alpha}e^{i\omega^\prime t(\alpha_2+u\alpha_3)}e^{i\omega t(1-\alpha_2-u\alpha_3)}\times{1\over2}\times\Big(-\frac{u v\cdot q}{24\pi^2 t^2}\Psi_{4;\pi}(\underline{\alpha})+\frac{u v \cdot q}{24\pi^2 t^2}\widetilde\Psi_{4;\pi}(\underline{\alpha})
\\ \nonumber &&+\frac{v \cdot q}{24\pi^2 t^2}\Phi_{4;\pi}(\underline{\alpha})-\frac{v\cdot q}{24\pi^2 t^2}\widetilde\Phi_{4;\pi}(\underline{\alpha})+\frac{v\cdot q}{24\pi^2 t^2}\Psi_{4;\pi}(\underline{\alpha})-\frac{v \cdot q}{24\pi^2 t^2}\widetilde\Psi_{4;\pi}(\underline{\alpha})\Big) \, .
\end{eqnarray}
The sum rule equations for the $\Xi_b^{\prime-}[{5\over2}^-]$ belonging to $[\mathbf{6}_F, 2 , 1, \lambda]$ are
\begin{eqnarray}
&&G_{\Xi_c^{\prime0}[{5\over2}^-]\to\Xi_c^{+}\pi^-}^D(\omega,\omega^{\prime})= \frac{g_{\Xi_c^{\prime0}[{5\over2}^-]\to\Xi_c^{+}\pi^- } f_{\Xi_c^{\prime0}[{5\over2}^-]}f_{\Xi_c^{+}}}{(\bar{\Lambda}_{\Xi_c^{\prime0}[{5\over2}^-]}-\omega^{\prime})(\bar{\Lambda}_{\Xi_c^{+}}-\omega)}
\\ \nonumber &=& \int_0^\infty dt \int_0^1 du e^{i(1-u)\omega^\prime t} e^{iu\omega t}\times 4\times \Big(\frac{3 f_{\pi} m_s u}{20\pi^2 t^2}\phi_{2;\pi}(u)+\frac{f_{\pi} m_{\pi}^2 u}{20(m_u+m_d)\pi^2}\phi_{3;\pi}^{\sigma}(u)
\\ \nonumber &&+\frac{3 f_{\pi} m_s u}{320\pi^2}\phi_{4;\pi}(u)+\frac{f_{\pi} u}{20}\langle\bar s s\rangle\phi_{2;\pi}(u)+\frac{f_{\pi} m_s m_{\pi}^2 u t^2}{480(m_u+m_d)}\langle\bar s s\rangle\phi_{3;\pi}^{\sigma}(u)+\frac{f_{\pi} u t^2}{320}\langle\bar s s\rangle\phi_{4;\pi}(u)
\\ \nonumber &&+\frac{f_{\pi} u t^2}{320}\langle g_s\bar s \sigma G s\rangle\phi_{2;\pi}(u)+\frac{f_{\pi} u t^4}{5120}\langle g_s\bar s \sigma G s\rangle\phi_{4;\pi}(u)\Big)
\\ \nonumber &-&\int_0^\infty dt \int_0^1 du \int \mathcal{D}\underline{\alpha}e^{i\omega^\prime t(\alpha_2+u\alpha_3)}e^{i\omega t(1-\alpha_2-u\alpha_3)}\times{1\over2}\times\Big(\frac{3f_{3\pi} u}{10\pi^2 t^2}\Phi_{3;\pi}(\underline{\alpha})-\frac{3f_{3\pi}}{10\pi^2 t^2}\Phi_{3;\pi}(\underline{\alpha})
\\ \nonumber &&=\frac{3i f_{3\pi}\alpha_3 u^2 v\cdot q}{10\pi^2 t}\Phi_{3;\pi}(\underline{\alpha})+\frac{3if_{3\pi}\alpha_2 u v\cdot q}{10\pi^2 t}\Phi_{3;\pi}(\underline{\alpha})-\frac{3i f_{3\pi} u v\cdot q}{10\pi^2 t}\Phi_{3;\pi}(\underline{\alpha})\Big) \, ,
\\
&&G_{\Xi_c^{\prime0}[{5\over2}^-]\to\Lambda_c^{+}K^-}^D(\omega,\omega^{\prime})= \frac{g_{\Xi_c^{\prime0}[{5\over2}^-]\to\Lambda_c^{+}K^- } f_{\Xi_c^{\prime0}[{5\over2}^-]}f_{\Lambda_c^{+}}}{(\bar{\Lambda}_{\Xi_c^{\prime0}[{5\over2}^-]}-\omega^{\prime})(\bar{\Lambda}_{\Lambda_c^{+}}-\omega)}
\\ \nonumber &=& \int_0^\infty dt \int_0^1 du e^{i(1-u)\omega^\prime t} e^{iu\omega t}\times 4\times \Big(\frac{f_{K} m_{K}^2 u}{20(m_u+m_s)\pi^2}\phi_{3;K}^{\sigma}(u)
\\ \nonumber &&+\frac{f_{K} u}{20}\langle\bar q q\rangle\phi_{2;K}(u)+\frac{f_{K} u t^2}{320}\langle\bar q q\rangle\phi_{4;K}(u)
\\ \nonumber &&+\frac{f_{K} u t^2}{320}\langle g_s\bar q \sigma G q\rangle\phi_{2;K}(u)+\frac{f_{K} u t^4}{5120}\langle g_s\bar q \sigma G q\rangle\phi_{4;K}(u)\Big)
\\ \nonumber &-&\int_0^\infty dt \int_0^1 du \int \mathcal{D}\underline{\alpha}e^{i\omega^\prime t(\alpha_2+u\alpha_3)}e^{i\omega t(1-\alpha_2-u\alpha_3)}\times{1\over2}\times\Big(\frac{3f_{3K} u}{10\pi^2 t^2}\Phi_{3;K}(\underline{\alpha})-\frac{3f_{3K}}{10\pi^2 t^2}\Phi_{3;K}(\underline{\alpha})
\\ \nonumber &&=\frac{3i f_{3K}\alpha_3 u^2 v\cdot q}{10\pi^2 t}\Phi_{3;K}(\underline{\alpha})+\frac{3if_{3K}\alpha_2 u v\cdot q}{10\pi^2 t}\Phi_{3;K}(\underline{\alpha})-\frac{3i f_{3K} u v\cdot q}{10\pi^2 t}\Phi_{3;K}(\underline{\alpha})\Big) \, ,
\\
&&G_{\Xi_c^{\prime0}[{5\over2}^-]\to\Xi_c^{\prime+}\pi^-}^D(\omega,\omega^{\prime})= \frac{g_{\Xi_c^{\prime0}[{5\over2}^-]\to\Xi_c^{\prime+}\pi^- } f_{\Xi_c^{\prime0}[{5\over2}^-]}f_{\Xi_c^{\prime+}}}{(\bar{\Lambda}_{\Xi_c^{\prime0}[{5\over2}^-]}-\omega^{\prime})(\bar{\Lambda}_{\Xi_c^{\prime+}}-\omega)}
\\ \nonumber &=& \int_0^\infty dt \int_0^1 du e^{i(1-u)\omega^\prime t} e^{iu\omega t}\times 4\times \Big(\frac{3 i f_{\pi} u}{10\pi^2 t^3}\phi_{2;\pi}(u)-\frac{i f_{\pi} m_s m_{\pi}^2 u}{40(m_u+m_d)\pi^2 t}\phi_{3;\pi}^{\sigma}(u)
\\ \nonumber &&+\frac{3 i f_{\pi} u}{160\pi^2 t}\phi_{4;\pi}(u)+\frac{i f_{\pi} m_s u t}{80}\langle\bar s s\rangle\phi_{2;\pi}(u)-\frac{i f_{\pi} m_{\pi}^2 u t}{120(m_u+m_d)}\langle\bar s s\rangle\phi_{3;\pi}^{\sigma}(u)+\frac{i f_{\pi} m_s u t^3}{1280}\langle\bar s s\rangle\phi_{4;\pi}(u)
\\ \nonumber &&-\frac{i f_{\pi} m_{\pi}^2 u t^3}{1920}\langle g_s \bar s \sigma G s\rangle\phi_{3;\pi}^{\sigma}(u)\Big)
\\ \nonumber &-&\int_0^\infty dt \int_0^1 du \int \mathcal{D}\underline{\alpha}e^{i\omega^\prime t(\alpha_2+u\alpha_3)}e^{i\omega t(1-\alpha_2-u\alpha_3)}\times{1\over2}\times\Big(\frac{3 f_{\pi} u}{20\pi^2 t^2 v\cdot q}\Psi_{4;\pi}(\underline{\alpha})-\frac{3f_{\pi} u}{20\pi t^2 v\cdot q}\widetilde\Psi_{4;\pi}(\underline{\alpha})
\\ \nonumber &&-\frac{3 f_{\pi}}{20\pi^2 t^2 v\cdot q}\Phi_{4;\pi}(\underline{\alpha})+\frac{3f_{\pi}}{20\pi^2 t^2 v\cdot q}\widetilde\Phi_{4;\pi}(\underline{\alpha})-\frac{3f_{\pi}}{20\pi^2 t^2 v\cdot q}\Psi_{4;\pi}(\underline{\alpha})+\frac{3 f_{\pi}}{20\pi^2 t^2 v\cdot q}\widetilde\Psi_{4;\pi}(\underline{\alpha})
\\ \nonumber &&+\frac{3 i f_{\pi}\alpha_3 u^2}{20\pi^2 t}\Phi_{4;\pi}(\underline{\alpha})+\frac{3 i f_{\pi}\alpha_2 u}{20\pi^2 t}\Phi_{4;\pi}(\underline{\alpha})+\frac{3if_{\pi}\alpha_3 u}{40\pi^2 t}\Phi_{4;\pi}(\underline{\alpha})+\frac{3if_{\pi}\alpha_3 u}{40\pi^2 }\widetilde\Phi_{4;\pi}(\underline{\alpha})
\\ \nonumber &&-\frac{3if_{\pi}u}{20\pi^2 t}\Phi_{4;\pi}(\underline{\alpha})+\frac{3if_{\pi}\alpha_2}{40\pi^2 t}\Phi_{4;\pi}(\underline{\alpha})+\frac{3i f_{\pi}\alpha_2}{40\pi^2 t}\widetilde\Phi_{4;\pi}(\underline{\alpha})-\frac{3if_{\pi}}{40\pi^2 t}\Phi_{4;\pi}(\underline{\alpha})-\frac{3if_{\pi}}{40\pi^2 t}\widetilde\Phi_{4;\pi}(\underline{\alpha})\Big) \, ,
\\
&&G_{\Xi_c^{\prime0}[{5\over2}^-]\to\Sigma_c^{\prime+}K^-}^D(\omega,\omega^{\prime})= \frac{g_{\Xi_c^{\prime0}[{5\over2}^-]\to\Sigma_c^{\prime+}K^- } f_{\Xi_c^{\prime0}[{5\over2}^-]}f_{\Sigma_c^{\prime+}}}{(\bar{\Lambda}_{\Xi_c^{\prime0}[{5\over2}^-]}-\omega^{\prime})(\bar{\Lambda}_{\Sigma_c^{\prime+}}-\omega)}
\\ \nonumber &=& \int_0^\infty dt \int_0^1 du e^{i(1-u)\omega^\prime t} e^{iu\omega t}\times 4\times \Big(\frac{3 i f_{K} u}{10\pi^2 t^3}\phi_{2;K}(u)+\frac{3 i f_{K} u}{160\pi^2 t}\phi_{4;K}(u)-\frac{i f_{K} m_{K}^2 u t}{120(m_u+m_s)}\langle\bar q q\rangle\phi_{3;K}^{\sigma}(u)
\\ \nonumber &&-\frac{i f_{K} m_{K}^2 u t^3}{1920}\langle g_s \bar q \sigma G q\rangle\phi_{3;K}^{\sigma}(u)\Big)
\\ \nonumber &-&\int_0^\infty dt \int_0^1 du \int \mathcal{D}\underline{\alpha}e^{i\omega^\prime t(\alpha_2+u\alpha_3)}e^{i\omega t(1-\alpha_2-u\alpha_3)}\times{1\over2}\times\Big(\frac{3 f_{K} u}{20\pi^2 t^2 v\cdot q}\Psi_{4;K}(\underline{\alpha})-\frac{3f_{K} u}{20\pi t^2 v\cdot q}\widetilde\Psi_{4;K}(\underline{\alpha})
\\ \nonumber &&-\frac{3 f_{K}}{20\pi^2 t^2 v\cdot q}\Phi_{4;K}(\underline{\alpha})+\frac{3f_{K}}{20\pi^2 t^2 v\cdot q}\widetilde\Phi_{4;K}(\underline{\alpha})-\frac{3f_{K}}{20\pi^2 t^2 v\cdot q}\Psi_{4;K}(\underline{\alpha})+\frac{3 f_{K}}{20\pi^2 t^2 v\cdot q}\widetilde\Psi_{4;K}(\underline{\alpha})
\\ \nonumber &&+\frac{3 i f_{K}\alpha_3 u^2}{20\pi^2 t}\Phi_{4;K}(\underline{\alpha})+\frac{3 i f_{K}\alpha_2 u}{20\pi^2 t}\Phi_{4;K}(\underline{\alpha})+\frac{3if_{K}\alpha_3 u}{40\pi^2 t}\Phi_{4;K}(\underline{\alpha})+\frac{3if_{K}\alpha_3 u}{40\pi^2 }\widetilde\Phi_{4;K}(\underline{\alpha})
\\ \nonumber &&-\frac{3if_{K}u}{20\pi^2 t}\Phi_{4;K}(\underline{\alpha})+\frac{3if_{K}\alpha_2}{40\pi^2 t}\Phi_{4;K}(\underline{\alpha})+\frac{3i f_{K}\alpha_2}{40\pi^2 t}\widetilde\Phi_{4;K}(\underline{\alpha})-\frac{3if_{K}}{40\pi^2 t}\Phi_{4;K}(\underline{\alpha})-\frac{3if_{K}}{40\pi^2 t}\widetilde\Phi_{4;K}(\underline{\alpha})\Big) \, ,
\\
&&G_{\Xi_c^{\prime0}[{5\over2}^-]\to\Xi_c^{*+}\pi^-}^D(\omega,\omega^{\prime})= \frac{g_{\Xi_c^{\prime0}[{5\over2}^-]\to\Xi_c^{*+}\pi^- } f_{\Xi_c^{\prime0}[{5\over2}^-]}f_{\Xi_c^{*+}}}{(\bar{\Lambda}_{\Xi_c^{\prime0}[{5\over2}^-]}-\omega^{\prime})(\bar{\Lambda}_{\Xi_c^{*+}}-\omega)}
\\ \nonumber &=& \int_0^\infty dt \int_0^1 du e^{i(1-u)\omega^\prime t} e^{iu\omega t}\times 4\times \Big(\frac{4if_{\pi}u}{15\pi^2 t^3}\phi_{2;\pi}(u)+\frac{if_{\pi} m_s m_{\pi}^2 u}{45(m_u+m_d)\pi^2 t}\phi_{3;\pi}^{\sigma}(u)
\\ \nonumber &&-\frac{if_{\pi}u}{60\pi^2 t}\phi_{4;\pi}(u)-\frac{i f_{\pi} m_s u t}{90}\langle\bar s s\rangle\phi_{2;\pi}(u)+\frac{i f_{\pi} m_{\pi}^2 u t}{135}\langle\bar s s\rangle\phi_{3;\pi}^{\sigma}(u)-\frac{i f_{\pi} m_s u t^3}{1440}\langle\bar s s\rangle\phi_{4;\pi}(u)
\\ \nonumber &&+\frac{i f_{\pi} m_{\pi}^2 u t^3}{2160}\langle g_s \bar s \sigma G s\rangle\phi_{3;\pi}^{\sigma}(u)\Big)
\\ \nonumber &-&\int_0^\infty dt \int_0^1 du \int \mathcal{D}\underline{\alpha}e^{i\omega^\prime t(\alpha_2+u\alpha_3)}e^{i\omega t(1-\alpha_2-u\alpha_3)}\times{1\over2}\times\Big(-\frac{2 f_{\pi} u}{15\pi^2 t^2 v\cdot q}\Psi_{4;\pi}(\underline{\alpha})+\frac{2 f_{\pi} u}{15 \pi^2 t^2 v\cdot q}\widetilde\Psi_{4;\pi}(\underline{\alpha})
\\ \nonumber &&+\frac{2 f_{\pi}}{15\pi^2 t^2 v\cdot q}\Phi_{4;\pi}(\underline{\alpha})-\frac{2 f_{\pi}}{15\pi^2 t^2 v\cdot q}\widetilde\Phi_{4;\pi}(\underline{\alpha})+\frac{2 f_{\pi}}{15\pi^2 t^2 v\cdot q}\Psi_{4;\pi}(\underline{\alpha})-\frac{2 f_{\pi}}{15\pi^2 t^2 v\cdot q}\widetilde\Psi_{4;\pi}(\underline{\alpha})
\\ \nonumber &&-\frac{i f_{\pi}\alpha_3 u^2}{15\pi^2 t}\Phi_{4;\pi}(\underline{\alpha})-\frac{2 f_{\pi}\alpha_2 u}{15\pi^2 t}\Phi_{4;\pi}(\underline{\alpha})-\frac{i f_{\pi}\alpha_3 u}{15\pi^2 t}\Phi_{4;\pi}(\underline{\alpha})-\frac{i f_{\pi}\alpha_3 u}{15\pi^2 t}\widetilde\Phi_{4;\pi}(\underline{\alpha})
\\ \nonumber &&+\frac{2 i f_{\pi} u}{15\pi^2 t}\Phi_{4;\pi}(\underline{\alpha})-\frac{i f_{\pi}\alpha_2}{15\pi^2 t}\Phi_{4;\pi}(\underline{\alpha})-\frac{i f_{\pi}\alpha_2}{15\pi^2 t}\widetilde\Phi_{4;\pi}(\underline{\alpha})+\frac{i f_{\pi}}{15\pi^2 t}\Phi_{4;\pi}(\underline{\alpha})+\frac{i f_{\pi}}{15\pi^2 t}\widetilde\Phi_{4;\pi}(\underline{\alpha})\Big) \, ,
\\
&&G_{\Xi_c^{\prime0}[{5\over2}^-]\to\Sigma_c^{*+}K^-}^D(\omega,\omega^{\prime})= \frac{g_{\Xi_c^{\prime0}[{5\over2}^-]\to\Sigma_c^{*+}K^- } f_{\Xi_c^{\prime0}[{5\over2}^-]}f_{\Sigma_c^{*+}}}{(\bar{\Lambda}_{\Xi_c^{\prime0}[{5\over2}^-]}-\omega^{\prime})(\bar{\Lambda}_{\Sigma_c^{*+}}-\omega)}
\\ \nonumber &=& \int_0^\infty dt \int_0^1 du e^{i(1-u)\omega^\prime t} e^{iu\omega t}\times 4\times \Big(\frac{4if_{K}u}{15\pi^2 t^3}\phi_{2;K}(u)-\frac{if_{K}u}{60\pi^2 t}\phi_{4;K}(u)+\frac{i f_{K} m_{\pi}^2 u t}{135}\langle\bar q q\rangle\phi_{3;K}^{\sigma}(u)
\\ \nonumber &&+\frac{i f_{K} m_{K}^2 u t^3}{2160}\langle g_s \bar q \sigma G q\rangle\phi_{3;K}^{\sigma}(u)\Big)
\\ \nonumber &-&\int_0^\infty dt \int_0^1 du \int \mathcal{D}\underline{\alpha}e^{i\omega^\prime t(\alpha_2+u\alpha_3)}e^{i\omega t(1-\alpha_2-u\alpha_3)}\times{1\over2}\times\Big(-\frac{2 f_{K} u}{15\pi^2 t^2 v\cdot q}\Psi_{4;K}(\underline{\alpha})+\frac{2 f_{K} u}{15 \pi^2 t^2 v\cdot q}\widetilde\Psi_{4;K}(\underline{\alpha})
\\ \nonumber &&+\frac{2 f_{K}}{15\pi^2 t^2 v\cdot q}\Phi_{4;K}(\underline{\alpha})-\frac{2 f_{K}}{15\pi^2 t^2 v\cdot q}\widetilde\Phi_{4;K}(\underline{\alpha})+\frac{2 f_{K}}{15\pi^2 t^2 v\cdot q}\Psi_{4;K}(\underline{\alpha})-\frac{2 f_{K}}{15\pi^2 t^2 v\cdot q}\widetilde\Psi_{4;K}(\underline{\alpha})
\\ \nonumber &&-\frac{i f_{K}\alpha_3 u^2}{15\pi^2 t}\Phi_{4;K}(\underline{\alpha})-\frac{2 f_{K}\alpha_2 u}{15\pi^2 t}\Phi_{4;K}(\underline{\alpha})-\frac{i f_{K}\alpha_3 u}{15\pi^2 t}\Phi_{4;K}(\underline{\alpha})-\frac{i f_{K}\alpha_3 u}{15\pi^2 t}\widetilde\Phi_{4;K}(\underline{\alpha})
\\ \nonumber &&+\frac{2 i f_{K} u}{15\pi^2 t}\Phi_{4;K}(\underline{\alpha})-\frac{i f_{K}\alpha_2}{15\pi^2 t}\Phi_{4;K}(\underline{\alpha})-\frac{i f_{K}\alpha_2}{15\pi^2 t}\widetilde\Phi_{4;K}(\underline{\alpha})+\frac{i f_{K}}{15\pi^2 t}\Phi_{4;K}(\underline{\alpha})+\frac{i f_{K}}{15\pi^2 t}\widetilde\Phi_{4;K}(\underline{\alpha})\Big) \, .
\end{eqnarray}
\end{widetext}

%
%%%%%%%%%%%%%%%%%%%%%%%%%%%%%%%%%%%%%%%%%%%%%%%%%%%%%%%%%%%%%%%%%%%%%%%%%%%%%%

%%%%%%%%%%%%%%%%%%%%%%%%%%%%%%%%%%%%%%%%%%%%%%%%%%%%%%%%%%%%%%%%%%%%%%%%%%%%%%
%

\end{document}